\documentclass[a4paper,oneside,reqno,11pt]{amsart}
\usepackage{amssymb}
\usepackage[foot]{amsaddr}
\usepackage{pdflscape}
\usepackage{url}
\usepackage{cite}
\usepackage{tikz}
\usetikzlibrary{graphs,backgrounds,quotes,positioning,arrows,matrix,shapes}
\usetikzlibrary{arrows.meta}
\usepackage{booktabs}
\usepackage{caption}
\usepackage{float}
\usepackage[bottom=3cm,top=3cm]{geometry}
\usepackage{hyperref}
\hypersetup{pdfpagelabels,pdftex,bookmarks,breaklinks}
\definecolor{darkblue}{RGB}{0,0,127} 
\definecolor{darkgreen}{RGB}{0,150,0}
\hypersetup{colorlinks, linkcolor=darkblue, citecolor=darkgreen, filecolor=red, urlcolor=blue}
\newtheorem{conjecture}{Conjecture}
\newcommand{\ea}[1]{\begin{align}#1\end{align}}
\newcommand{\nn}{\nonumber \\}
\newcommand{\vpu}[1]{^{\vphantom{#1}}}
\newcommand{\bc}{\begin{center}}
\newcommand{\ec}{\end{center}}
\newcommand{\bmt}{\begin{pmatrix}}
\newcommand{\emt}{\end{pmatrix}}

\newcommand{\mbf}[1]{\mathbf{#1}}
\newcommand{\mcl}[1]{\mathcal{#1}}
\newcommand{\fd}[1]{\mathbb{#1}}
\newcommand{\la}{\langle}
\newcommand{\ra}{\rangle}
\newcommand{\btenv}[1]{\parbox[t]{4.6 cm}{ \centering #1}}
\DeclareMathOperator{\GCD}{GCD}
\DeclareMathOperator{\Tr}{Tr}
\DeclareMathOperator{\Det}{Det}
\DeclareMathOperator{\gal}{Gal}
\DeclareMathOperator{\C}{C}
\DeclareMathOperator{\EC}{EC}

\DeclareMathOperator{\SL}{SL}
\DeclareMathOperator{\ESL}{ESL}
\DeclareMathOperator{\GL}{GL}
\DeclareMathOperator{\re}{Re}

\title[Constructing exact SICs from numerical solutions]{Constructing exact symmetric informationally complete measurements from numerical solutions}

\author{$^*$Marcus Appleby}
\email{marcus.appleby@sydney.edu.au}
\author{$^\dagger$Tuan-Yow Chien}
\email{tuan@math.auckland.ac.nz}
\author{$^{*,\ddagger}$Steven Flammia}
\email{steven.flammia@sydney.edu.au}
\author{$^\dagger$Shayne Waldron}
\email{waldron@math.auckland.ac.nz}

\address{$^*$Centre for Engineered Quantum Systems, School of Physics, University of Sydney, Sydney, Australia}
\address{$^\dagger$Department of Mathematics, University of Auckland, Auckland, New Zealand}
\address{$^\ddagger$Center for Theoretical Physics, Massachusetts Institute of Technology, Cambridge, USA}

\begin{document}

\begin{abstract}
Recently, several intriguing conjectures have been proposed connecting symmetric informationally complete quantum measurements (SIC POVMs, or SICs) and algebraic number theory.
These conjectures relate the SICs and their minimal defining algebraic number field. 
Testing or sharpening these conjectures requires that the SICs are expressed exactly, rather than as numerical approximations. 
While many exact solutions of SICs have been constructed previously using Gr{\"{o}}bner bases, this method has probably been taken as far as is possible with current computer technology (except in special cases, where there are additional symmetries). 
Here we describe a method for converting high-precision numerical solutions into exact ones using an integer relation algorithm in conjunction with the Galois symmetries of a SIC. 
Using this method we have calculated 69 new exact solutions, including 9 new dimensions where previously only numerical solutions were known, which more than triples the number of known exact solutions. 
In some cases the solutions require number fields with degrees as high as 12,288. 
We use these solutions to confirm that they obey the number-theoretic conjectures and we address two questions suggested by the previous work.
\end{abstract}

\maketitle

\section{Introduction}
Symmetric informationally complete quantum measurements (SIC POVMs as they are often called, or SICs as we will call them in this paper) are collections of $d^2$ equiangular lines in in a $d$-dimensional complex vector space. 
They were originally introduced by Zauner\cite{Zauner:1999} and Renes \emph{et al}~\cite{Renes:2004}, and they have many applications to quantum information~~\cite{Fuchs:2003,Rehacek:2004,Englert:2004,Scott:2006,Durt:2008,Zhu:2011,Zhu:2016a}, as well as playing a central role in the QBist approach to the interpretation of quantum mechanics~\cite{Fuchs:2013}. 
They also have applications to classical signal processing~\cite{Howard:2006,Malikiosis:2016}. 
SICs have been calculated numerically\cite{Renes:2004,Scott:2010a,Scott:2017,Fuchs:2017,Grassl:2017,Grassl:2018} in every dimension up to 151, and for a handful of other dimensions up to $1155$. 
Exact solutions\cite{Zauner:1999,Appleby:2005b,Scott:2010a,Appleby:2012,Appleby:2014,Grassl:2017,Grassl:2018} have been calculated in dimensions $2$--$16$, $19$, $24$, $28$, $35$, $48$, $120$,  $124$, $323$. 
This encourages the conjecture that SICs exist in every finite dimension, but a proof continues to elude us. 
The SICs in dimensions 2 and 3, together with the Hoggar lines~\cite{Hoggar:1998} in dimension 8, are sometimes called sporadic SICs, on the grounds that they have a number of special properties~\cite{Zhu:2015a,Stacey:2016}. 
In this paper we exclude them from consideration.  In the following the term ``SIC'' will therefore always mean ``non-sporadic SIC''.  In particular, we will always assume without comment that the dimension is greater than 3. 

In physics a numerical solution to a system of equations is usually perfectly adequate, especially when, as here, it  is known to 35 digits of precision or more.  However, as we  discuss below, SICs have important features (important for the applications as well as from a pure mathematical point of view) which are only accessible if one has an exact solution in hand.  Here we face the difficulty that, aside from a few low dimensional cases where a hand calculation is possible, all previously known exact solutions were obtained using a Gr{\"{o}}bner basis.  Moreover every solution in dimensions higher than 16 relied on the existence of additional symmetries which are only present in special cases.  The great majority of SICs do not have such symmetries, and for these the Gr{\"{o}}bner basis method has probably been pushed about as far as it is possible to go without massive increases in computer speed and memory.  We therefore need a new method for calculating exact SICs.  In this paper we describe such a method.  The idea is to start with a high-precision numerical solution (much easier to calculate) and then convert it into an exact solution.

The method relies on, and is motivated by some recently discovered connections between SICs and algebraic number theory~\cite{Marcus:1977,Cohn:1978}.  Dimension 3 excepted, the standard-basis matrix elements of every known exact  fiducial projector are algebraic numbers.  More than that, they are expressible in radicals (i.e.\ the components can be built up from the integers using the standard arithmetical operations together with the operation of taking roots), meaning that the associated Galois group is solvable.  It turns out that the number fields they generate  have some remarkable properties~\cite{Appleby:2013,Appleby:2016,Appleby:2017a}. We outline the most salient points of this number-theoretic connection here assuming some familiarity with the basic concepts underlying SICs; however, in section~\ref{sec:prelim} below we provide all of the necessary definitions and background. 

Let $\Pi$ be a SIC fiducial projector in dimension $d\ge 4$, and let $\fd{E}=\fd{Q}(\Pi,\tau)$ be the  field generated over the rationals by the standard-basis components of $\Pi$  together with $\tau = -e^{\frac{\pi i}{d}}$.  We refer to $\fd{E}$ as the SIC field.  It is easily seen that $\fd{E}$ only depends on the $\EC(d)$ orbit to which $\Pi$ belongs (where $\EC(d)$ is the extended Clifford group in dimension $d$).  Let $\fd{K} = \fd{Q}(\sqrt{D})$ where $D$ is the square-free part of $(d-3)(d+1)$.  Then one finds, in every case that has been calculated,
\begin{itemize}
\item $\fd{Q} \triangleleft \fd{K} \triangleleft \fd{E}$.
\item $\fd{Q} \triangleleft \fd{E}$.
\item The Galois group $\gal(\fd{E}/\fd{K})$ is  Abelian.
\end{itemize}
(where the notation $\fd{F}_1 \triangleleft \fd{F}_2$ means that $\fd{F}_2$ is a normal extension of $\fd{F}_1$).  
In each dimension for which the full set of exact SICs has been calculated there is exactly one SIC  field of minimal degree; every other SIC field  being a low-degree extension of that.  In ref.~\cite{Appleby:2016} it was shown that, in every known case,  the minimal field is a very special kind of field extension:  Namely, the ray class field over $\fd{K}$ with conductor $d'$ and ramification at both infinite places (where $d'=d$ if $d$ is odd, and $2d$ if $d$ is even).

If SICs existed in every finite dimension, and if the above statements were generally true, it would be a fact of some interest to algebraic number theorists.  Ray class fields are used to classify fields having an Abelian Galois group over some given base-field.  The Kronecker-Weber theorem states that the ray class fields over $\fd{Q}$ are precisely the fields $\fd{Q}(\omega)$, where $\omega$ is a complex root of unity.  Kronecker further conjectured, and it was subsequently proved, that the ray class fields over $\fd{Q}(i\sqrt{n})$, for $n$ a positive integer, are generated by the coordinates of certain distinguished points on an elliptic curve.  Hilbert's $12^{\rm{th}}$ problem, still unsolved, asks for the generalization of these results; the obvious place to start being ray class fields over $\fd{Q}(\sqrt{n})$, for $n$ a positive integer---i.e., fields of precisely the kind that SICs generate (in the handful of cases we have been able to calculate).  This does not mean that solving the SIC problem is  equivalent to solving Hilbert's $12^{\rm{th}}$ problem.  In the first place  SICs only give us \emph{some} of the ray class fields over a given $\fd{Q}(\sqrt{n})$.  In the second place a solution to the SIC problem would not necessarily give us the analogues of the exponential and elliptic functions that Hilbert was asking for.  Against that, it was shown in ref.~\cite{Appleby:2016} that, for each square-free positive integer $n$, there is an infinite sequence of dimensions for which $D=n$.  So although a solution to the SIC problem would not give us the full set of ray class fields over a given $\fd{Q}(\sqrt{n})$, it might give us an infinite subset.  Moreover, the SIC problem may conceivably reduce to proving a set of special-function identities.  If so that would give us the  functions Hilbert wanted.  In short, it seems fair to say that a constructive solution to the SIC problem might be a significant step in the direction of solving Hilbert's $12^{\rm{th}}$ problem for real quadratic fields.  

We have stressed the potential relevance of SICs to a major unsolved problem in algebraic number theory.  The reverse is also true:  the number-theoretic aspects give important insight into SIC geometry, as appears from ref.~\cite{Appleby:2017b} which describes a series of significant geometrical results, none of which would have been suspected without the clue from number theory, and which cannot be properly understood in isolation from the number theory.  Two of the authors are actively engaged in pursuing these connections between number theory and geometry further.

For many purposes a high-precision approximate SIC is completely satisfactory.  However, if one wants to investigate the connections with number theory exact solutions are essential. As we discussed above, the Gr{\"{o}}bner basis method, used to obtain most of the previously known exact solutions, is extremely demanding computationally and is probably not capable of being taken much further on existing computer hardware.   Fortunately, number theory, besides creating the demand, also supplies the means of satisfying it.  That is, one can exploit the conjectured number-theoretic properties of a SIC to bump the high precision approximate SICs in refs~\cite{Scott:2010a,Scott:2017} up to infinite precision.  Success in this enterprise incidentally provides additional evidence in support of the conjectures on which the method is based.

The essential idea is as follows.  Suppose one is given a real number $a$ specified to some finite degree of precision, and suppose one knows that $a$ is an approximation to an exact real number $a_{\rm{e}}$ in a specified algebraic number field $\fd{F}$.  Let $b_1, \dots, b_n$ be a basis for $\fd{F}$ over the rationals.  Then $a_{\rm{e}} = q_1 b_1 + \dots + q_n b_n$ for some set of rational numbers $q_j$.  Multiplying through by the LCM of the $q_j$ we deduce
\ea{
m_0 a - \sum_{j=1}^n m_j b_j \approx 0
\label{eq:IRAintro}
}
for some set of integers $m_0, m_1, \dots m_n$.  If we can calculate the $m_j$ we will have managed to find an exact number which $a$ approximates. 
With high enough precision and computational power, these particular integers $m_j$ will satisfy $q_j = m_j/m_0$, and we will have recovered $a_{\rm{e}}$, the specific exact number we're after. 
This procedure can be done using an integer-relation algorithm~\cite{Bailey:2007}, such as the PSLQ algorithm~\cite{Ferguson:1999}.

A few remarks are in order.  In the first place, there are infinitely many sequences $m_j$ satisfying Eq.~\eqref{eq:IRAintro} to the specified degree of accuracy, and there is no guarantee that the algorithm will return  the one we want.  Basically, the algorithm is  a systematic guessing procedure.  However, for our purposes that is good enough, since we  can verify the guess (by checking that the final result really is an exact SIC fiducial).

In the second place, one does not expect to get more information out than one initially puts in.  Suppose $a$ is given to $r$ digits of precision.  Then if one does not want to generate spurious results one needs $r$ to be larger than a number $\sim s(n+1)$, where $s$ is the maximum number of digits in the integers $m_j$.  In practice~\cite{Bailey:2007} one needs $r$ to be 10--15\% larger than $s(n+1)$.  The problem is, of course, that one does not initially know $s$.  As a practical rule of thumb we therefore proceed by repeatedly running the algorithm at successively higher levels of precision until the result is stable.  Again, the justification for this procedure is the fact that the end result demonstrably is an exact SIC fiducial.

In the third place, there is the problem that we do not always know the  field $\fd{F}$.  As discussed above, we have a conjecture regarding the minimal SIC field in each dimension.  However, many fiducials lie in an extension of this field for which we currently have no conjecture.  A further problem is that in the cases of interest the SIC field is often of degree $10^3$ or more.  For the reasons discussed above this means that we need to work to very high precision, and the calculations are correspondingly slow.  For the calculations in this paper we therefore exploited a conjecture in ref.~\cite{Appleby:2013}, which implies that the coefficients of suitably chosen polynomials lie in a much smaller  field, which is easily inferred from the numerical data, and for which the calculations are much faster.

Scott-Grassl~\cite{Scott:2010a} calculated 179 numerical fiducials in dimensions  $4$--$50$, and  argued that with high probability their list is complete, in the sense that it includes exactly one representative of every orbit of unitarily equivalent SICs in these dimensions.  Scott~\cite{Scott:2017} calculated a further 323 numerical fiducials in dimensions $51$--$90$, restricting the search to fiducials having an $F_z$ or $F_a$ symmetry.  He states that it is ``likely" that this list is complete (by contrast, he is ``confident'' that the list in ref.~\cite{Scott:2010a} is complete).  Another 76 numerical fiducials have been calculated~\cite{Scott:2017,Fuchs:2017,Grassl:2017,Grassl:2018} in dimensions $91$--$121$ and for a handful of additional dimensions up to  $1155$, but with no claim as to completeness.  This gives us a total of $578$ unitarily inequivalent numerical fiducials for $d\ge 4$.   By contrast, there are only $30$ orbits  for which an exact solution has been calculated~\cite{Zauner:1999,Appleby:2005b,Scott:2010a,Appleby:2012,Appleby:2014,Grassl:2017,Grassl:2018}  with $d\ge 4$.  
We have calculated exact representatives for a further $69$ orbits, and of these orbits we provide a complete analysis for $52$ of them. 
This means that we now have exact representatives for $96$ total orbits, more than half of the orbits in dimensions between $4$ and $50$.  
We would have liked to calculate exact representatives and provide a complete analysis for every orbit for which a numerical fiducial is known, but unfortunately our method is not completely automated and is still too time consuming for us to justify this additional effort.  
In choosing which fiducials to calculate we were guided by two considerations.  
In the first place the conjectures in ref.~\cite{Appleby:2013}, as was there noted, cannot be true of the type-$a$ orbits $21e$, $30d$, $39ghij$, $48e$ (where we employ the Scott-Grassl~\cite{Scott:2010a} labeling scheme).  
We have calculated exact representatives of these orbits in order to find modified conjectures which stand a chance of holding for every type-$a$ orbit.  
In the second place, we noted above that the minimal SIC field in each dimension seems always to be the ray class field over $\fd{Q}(\sqrt{D})$ with conductor $d'$.  
This raises the question, what can be said of the \emph{non}-minimal fields in each dimension.  
We therefore set out to find a \emph{complete} set of SICs in each dimension.  
Thus, Scott and Grassl only give an exact representative for one of the orbits in dimensions $15$, $19$, $24$, $28$, $35$ and $48$.  
We have found exact representatives for all the other orbits in these dimensions.  
In particular we have found full sets of exact solutions for dimensions $35$ and $39$---the two dimensions less than $50$ for which the number of distinct orbits is greatest. 
This information may help us to find a conjecture similar to the ray-class field conjecture applying to the non-minimal SIC fields.  

The complete list of dimensions for which we provide exact solutions, and where previously only a numerical solution was known, is $17, 18, 20, 21, 30, 31, 37, 39, 43$. 
From this list, we have neglected to provide a complete analysis for $31, 37, 43$, although our methods could also be applied there by a sufficiently motivated individual. 

The plan of the paper is as follows.  Section~\ref{sec:prelim} is devoted to necessary preliminaries.  Mostly this material is in refs.~\cite{Appleby:2005b,Scott:2010a,Appleby:2013,Appleby:2016,Appleby:2017a}.  
Some of it, however, is new.  
Section~\ref{sec:typea} is devoted to the type-$a$ orbits $21e$, $30d$, $39ghij$, $48e$ mentioned in the last paragraph.  
We show how the conjectures  in ref.~\cite{Appleby:2013}  naturally generalize to these orbits also.  
Section~\ref{sec:FieldsMultiplets}  concerns the relationship between the SIC fields in a given dimension.  
Scott and Grassl~\cite{Scott:2010a} find at most three exact solutions in a given dimension, and no more than two distinct SIC fields.  
With the new solutions calculated here we now have up to ten exact solutions for a given dimension, and up to seven distinct SIC fields.  
In section~\ref{sec:FieldsMultiplets} we summarize this information.  In section~\ref{sec:calc} we describe the method used to calculate the new solutions.  
Finally, in Appendix~\ref{sec:results} we present our results.  
Many of the fields we calculate are much higher degree\footnote{Calculating with fields this enormous leads to numerous technical difficulties.  For instance, we estimated that it would take many months to directly check the irreducibility of the successive polynomials used to define some of the field towers (by trying to factor them in \emph{Magma}), and we therefore had recourse to an  indirect argument based on the subfield structure.  Again, when calculating with such fields it is essential to devise ways of minimizing the number of distinct arithmetical operations (especially the number of divisions) if one does not want to devote months or even years of CPU time to the task. One also needs to give careful thought to the amount of RAM taken up by some of the intermediate results.} than the fields calculated in ref.~\cite{Scott:2010a}, and this creates some presentational difficulties.  
For instance the print-out for exact fiducial $48a$ occupies almost a thousand A4 pages (font size 9 and narrow margins).  
There can therefore be no question of presenting the fiducials themselves in the appendix.  
Instead we have made the fiducials available online~\cite{Exact:2016}, and confined ourselves here to a description of the fields and Galois groups.

\section{Preliminaries}
\label{sec:prelim}
The purpose of this section is to fix notation, and to summarize some relevant results from refs.~\cite{Appleby:2005b,Scott:2010a,Scott:2017,Appleby:2013,Appleby:2016,Appleby:2017a}. Some of the statements in this section are  proven facts  while others are  observations only known to be true for the SICs which have actually been calculated.  The latter will always be clearly identified as such.  As explained in the Introduction we assume without comment that the SICs with which we deal are non-sporadic in the sense of ref.~\cite{Stacey:2016}.  In particular it will always be assumed that the dimension $d\ge 4$, and that the SICs with which we deal are covariant with respect to the Weyl-Heisenberg (WH) group.

We first define the WH group.  Let $\mcl{H}$ be a $d$-dimensional Hilbert space, and let $|0\ra, \dots |d-1\ra$ be the standard basis. Let $X$, $Z$ by the operators which act according to
\ea{
X|r\ra &= | r+1\ra, &  Z |r\ra &= \omega^r |r\ra,
}
where $\omega = e^{2\pi i /d}$ and addition of indices is \emph{mod} $d$.  For each $\mbf{p} = \left(\begin{smallmatrix} p_1 \\ p_2 \end{smallmatrix}\right) \in \mathbb{Z}^2$ define the displacement operator
\ea{
D_{\mbf{p}} &= \tau^{p_1p_2} X^{p_1} Z^{p_2},
}
where $\tau= - e^{\frac{\pi i }{d}}$.  The WH group then consists of all operators of the form $\tau^r D_{\mbf{p}}$, for $r$ an integer.
 
The displacement operators constitute a nice unitary error basis~\cite{Knill:1996,Klappenecker:2002}.  In particular, an arbitrary operator $A$ has the expansion
\ea{
A &= \sum_{p_1,p_2 = 0}^{d-1} A_{\mbf{p}} D_{\mbf{p}}, &  A_{\mbf{p}} &= \frac{1}{d}\Tr(D^{\dagger}_{\mbf{p}} A).
\label{eq:DpBsis}
}

We next define the Clifford and extended Clifford groups.
Let $d' = d$ (respectively $d'=2d$) if $d$ is odd (respectively even), and define the extended symplectic group $\ESL(2,\fd{Z}/d'\fd{Z})$  to consist  of all $2\times 2$ matrices with entries in $\fd{Z}/d'\fd{Z}$ and determinant $\pm 1$.  The symplectic group $\SL(2,\fd{Z}/d'\fd{Z})$ is the subgroup consisting of matrices with determinant $+1$.  For each $F \in \ESL(2,\fd{Z}/d'\fd{Z})$ there exists an operator $U_F$ on $\mcl{H}_d$, unique up to an overall phase, such that
\ea{
U\vpu{\dagger}_F D_{\mbf{p}} U^{\dagger}_F &= D_{F\mbf{p}}.
}
The operator $U_F$ is a unitary if $\Det F=1$ and an anti-unitary if $\Det F = -1$. We refer to it as a symplectic unitary in the first case and an anti-symplectic anti-unitary  in the second.  For all $F,G \in \ESL(2,\fd{Z}/d'\fd{Z})$
\ea{
U_F U_G \dot{=} U_{FG}
}
where $\dot{=}$ means ``equal up to a phase'' (so the map
 $F\to U_F$ is a projective representation of $\ESL(2,\fd{Z}/d'\fd{Z})$). The  Clifford group $\C(d)$ (respectively extended Clifford  group $\EC(d)$) consists of all operators of the form $e^{i\theta} D_{\mbf{p}} U_F$ with $F\in \SL(2,\fd{Z}/d' \fd{Z})$ (respectively $F \in \ESL(2,\fd{Z}/d'\fd{Z}$) and $e^{i\theta}$ an arbitrary phase.  
 
There are~\cite{Appleby:2005b,Appleby:2009} explicit formulae for the operators $U_F$.  If the symplectic matrix
\ea{
F = \bmt \alpha & \beta \\ \gamma & \delta \emt
}
is such that $\GCD(\beta,d') = 1$ the associated unitary is given by
\ea{
U_F &= \frac{e^{i\theta}}{\sqrt{d}} \sum_{r,s=0}^{d-1} \tau^{\beta^{-1}(\delta r^2 - 2 r s + \alpha s^2)} |r\ra \la s|
\label{eq:UFformula}
}
where $e^{i\theta}$ is an arbitrary phase, and $\beta^{-1}$ is the multiplicative inverse of $\beta$ considered as an element of $\fd{Z}/d'\fd{Z}$.  Matrices in $\SL(2,\fd{Z}/d'\fd{Z})$ not satisfying the condition $\GCD(\beta,d') = 1$ can always be written as a product of two  matrices which do satisfy it. An antisymplectic matrix $F$ can be written as $F' J$ where $F' \in \SL(2,\fd{Z}/d'\fd{Z})$ and 
\ea{
J &= \bmt 1 & 0 \\ 0 & -1 \emt.
}
The anti-unitary $U_J$ acts by complex conjugation in the standard basis.  Using these facts we can calculate  $U_F$ for arbitrary $F\in \ESL(2,\fd{Z}/d'\fd{Z})$.


By definition a WH-covariant SIC consists of the $d^2$ operators $D\vpu{\dagger}_{\mbf{p}} \Pi D^{\dagger}_{\mbf{p}}$ where 
$\Pi$ is a rank-1 projector with the property
\ea{
\Tr(\Pi D_{\mbf{p}}) &=
\begin{cases}
 1  \qquad & \text{if $\mbf{p} = \boldsymbol{0} \mod d$}
\\
 \frac{e^{i\theta_{\mbf{p}}}}{\sqrt{d+1}} \qquad & \text{otherwise}
\end{cases}
}
for some set of phases $e^{i\theta_{\mbf{p}}}$.  We refer to $\Pi$ as the fiducial, to the quantities $\Tr(\Pi D_{\mbf{p}})$ as the overlaps, and to the numbers $e^{i\theta_{\mbf{p}}}$ as the overlap phases.  It can be seen from Eq.~\eqref{eq:DpBsis} that the overlap phases completely determine the fiducial.  

\vspace{0.1 cm}

 \noindent \textbf{Observation 1.} In every case which has been calculated, the overlap phases are  units in the field $\fd{E}(\sqrt{d+1})$  generating a subgroup of the full unit group whose properties are described in ref.~\cite{Appleby:2016}.  

\vspace{0.1 cm}

\noindent If $\Pi$ is a SIC fiducial projector then so is $U\Pi U^{\dagger}$ for every $U\in \EC(d)$.  Consequently the fiducial projectors split into a collection of $\EC(d)$ orbits.  
Scott and Grassl~\cite{Scott:2010a} have, with high probability, identified every $\EC(d)$ orbit for $d\le 50$ and Scott~\cite{Scott:2017} has extended the calculations to higher dimensions. 

\vspace{0.1 cm}

\noindent \textbf{Observation 2.} At the time of writing it appears that there are only finitely many $\EC(d)$ orbits in every dimension  $d\le 90$.  

\vspace{0.1 cm}

We define the stability group of a SIC fiducial $\Pi$ to be the set of all $U\in \EC(d)$ such that $U\Pi U^{\dagger} = \Pi$.  

\vspace{0.1 cm}

\noindent \textbf{Observation 3.} In every known case the stability group contains a unitary of the form $D_{\mbf{p}} U_F$, where $\Tr(F) = -1 \mod d$.

\vspace{0.1 cm}

\noindent Such a unitary is necessarily order 3, up to a phase.  We refer to it as canonical order 3. 

It is convenient to choose standard canonical order 3 unitaries.  If $d\neq 3$ or $6 \mod 9$ every canonical order 3 unitary is conjugate modulo a phase to $U_{F_z}$, where
\ea{
F_{z} &= \bmt 0 & d-1 \\ d+1 & d-1 \emt .
}
We refer to $F_z$ as the Zauner matrix\footnote{We define the Zauner matrix this way to secure consistency with ref.~\cite{Scott:2010a}.  The Zauner matrix is often defined to be $F^4_z  = \left(\begin{smallmatrix} 0 & -1 \\ 1 & -1 \end{smallmatrix}\right)$ which is order 3, unlike $F_z$ which is order 6 when $d$ is even.  This makes no difference at the level of the unitaries since $U_{F_z}$ and $U_{F^4_z}$ are equal modulo a phase.}.

\vspace{0.1 in}

\noindent \textbf{Observation 4.} In the cases where it is likely that we know the full set of $\EC(d)$ orbits it is found that if $d=6 \mod 9$ the additional conjugacy class never gives rise to SICs, while if $d=3\mod 9$ it always does.

\vspace{0.1 in}

\noindent  When $d=3\mod 9$ we choose~\cite{Scott:2010a}, as representative of the additional conjugacy class  $U_{F_a}$, where
\ea{
F_{a} &= \bmt 1 & d+3 \\ \frac{4d-3}{3} & d-2 \emt .
}
We say that an orbit is type-$z$ (respectively type-$a$) if it contains fiducials stabilized by $U_{F_z}$ (respectively $U_{F_a}$).

We say that a subgroup of $\EC(d)$ is displacement-free if it consists entirely of operators of the form $e^{i\theta} U_F$, with $F\in \ESL(2,\fd{Z}/d'\fd{Z})$.  Following  refs.~\cite{Appleby:2016,Appleby:2017a} we shall say that a fiducial is centred if its stability group (a) contains a canonical order 3 unitary, and (b) is displacement free (in ref.~\cite{Appleby:2013} such fiducials were called simple). 

\vspace{0.1 cm}

\noindent \textbf{Observation 5.}   Every known $\EC(d)$ orbit of SIC fiducials  contains at least one centred fiducial.

\vspace{0.1 cm}

\noindent Given a centred fiducial $\Pi$ we define $S_0(\Pi)$ to consist of all $F\in \ESL(2,\fd{Z}/d'\fd{Z})$ such that $U_F$ is in the stability group of $\Pi$. 
 
We next consider the Galois symmetries of the Clifford unitaries and SIC projectors.  We  say that an operator $A$ is in (respectively generates) a field $\fd{F}$ as shorthand for the statement that its standard basis matrix elements are in (respectively generate) $\fd{F}$.  Similarly we  write $\fd{F}(A)$ to mean $\fd{F}\left( \bigcup_{r,s=0}^d \la r | A | s\ra \right)$.  It is immediate that the displacement operators generate the cyclotomic field $\fd{Q}(\tau)$.  The same is true of the symplectic unitaries, provided the phase in  Eq.~\eqref{eq:UFformula} is chosen appropriately. This was shown in  ref.~\cite{Appleby:2015a} for the case of prime dimensions using an argument based on Gauss sums.  The argument is easily generalized to show, for arbitrary $d$, that if one chooses 
\ea{
e^{i\theta} &=
\begin{cases}
1 \qquad & d = 1 \mod 4
\\
i \qquad & d= 3 \mod 4
\\
e^{\frac{\pi i}{4}} \qquad & d= 0 \mod 2
\end{cases}
}
then $e^{i\theta}/\sqrt{d}$, and consequently $U_F$ is in $\fd{Q}(\tau)$ for all $F\in \SL(2,\fd{Z}/d'\fd{Z})$.  In the sequel we shall always assume that this choice of phase has been made.
 
Now let $\Pi$ be any SIC fiducial projector and define $\fd{E} = \fd{Q}(\Pi, \tau)$.  

\vspace{0.1 cm}

\noindent \textbf{Observation 6.}   In every known case the extension $\fd{E}/\fd{Q}$ is finite degree and normal. In particular $\gal(\fd{E}/\fd{Q})$ contains complex conjugation implying $\Pi^{*} \in \fd{E}$.

\vspace{0.1 cm}

\noindent Observation $6$ means that in every known case $\fd{E}$ is the same for every  fiducial projector on the same $\EC(d)$ orbit.  We refer to it as the SIC-field for that orbit\footnote{Note that in ref.~\cite{Appleby:2013} the extension $\bar{\fd{E}} = \fd{E}(\sqrt{d})$ was also introduced, to ensure that the Clifford unitaries would be in the field.  The considerations in the previous paragraph show that this was unnecessary.}.

\vspace{0.1 cm}

\noindent \textbf{Observation 7.} Let $\Pi$ be a centred fiducial, and let $g_c$ be complex conjugation.   In every known case $\fd{E}$  breaks into the tower
\ea{
\fd{Q} \vartriangleleft \fd{E}_c \vartriangleleft \fd{E}_0 \vartriangleleft \fd{E}_1 \vartriangleleft \fd{E}
\label{eq:tower}
}
where
\begin{enumerate}
\item The notation $\fd{F} \vartriangleleft \fd{G}$ means ``$\fd{G}$ is a normal extension of $\fd{F}$''.
\item \label{ls:ecDef}  $\fd{E}_c$ is the fixed field of the centralizer of $g_c$.
\item \label{ls:ecTermsD} $\fd{E}_c = \fd{Q}(\sqrt{D})$, where $D$ is the square-free part of $(d-3)(d+1)$.
\item \label{ls:centgc} $\gal(\fd{E}/\fd{E}_c)$ is Abelian,
\item For all $g\in \gal(\fd{E}/\fd{Q})$
 \begin{enumerate}
 \item  $g$ fixes $\fd{E}_c$ if and only if $g(\Pi)$ is a SIC fiducial projector ,
\item $g$ fixes $\fd{E}_0$ if and only if $g(\Pi)$ is a fiducial on the same $\EC(d)$ orbit as $\Pi$.
\end{enumerate}
\item $\fd{E}=\fd{E}_1(i\sqrt{d'})$. In particular $\fd{E}$ is a degree 2 extension of $\fd{E}_1$.
\item \label{ls:gbar} Let $\bar{g}_1$ be the non-trivial element of $\gal(\fd{E}/\fd{E}_1)$ and let $g$ be any element of $\gal(\fd{E}/\fd{Q})$ which does not fix $\fd{Q}(\sqrt{D})$.  Then
\ea{
\bar{g}_1 &= g g_{\rm{c}} g^{-1}.
\label{eq:gcgb1}
}
\item  For all $\mbf{p}$ and some $\mbf{k}_{\Pi}\in (\fd{Z}/d'\fd{Z})^2$
\ea{
\bar{g}_1 \bigl(\Tr(D_{\mbf{p}} \Pi)\bigr) = \omega^{\la \mbf{k}_{\Pi} , \mbf{p}\ra } \Tr(D_{\mbf{p}} \Pi).
\label{eq:E1fixcond}
} 
\item $\fd{E}_c$, $\fd{E}_0$, $\fd{E}$ are normal over $\fd{Q}$, but $\fd{E}_1$ is not.
\end{enumerate}
 Note  that it follows from item (\ref{ls:gbar}) that $\bar{g}_1(\tau) = \tau^{-1}$.

\vspace{0.1 cm}

We  say that  a fiducial to which Observation 7 applies is strongly centred if it is centred and $\mbf{k}_{\Pi}= 0$, so that the overlaps are all in $\fd{E}_1$.  If $d\neq 0 \mod 3$ centred fiducials are automatically strongly centred\cite{Appleby:2013}.  If $d=0 \mod 3$ the concepts are not equivalent.  However

\vspace{0.1 cm}

\noindent \textbf{Observation 8.}   For every $\EC(d)$ orbit for which an exact solution is known there exist strongly centred fiducials whose stability group includes $U_{F_z}$ (for a type-$z$ orbit) or $U_{F_a}$ (for a type-$a$ orbit)\footnote{This fact was not appreciated at the time ref.~\cite{Appleby:2013} was written.  Consequently, the representatives of $\EC(d)$ orbits  $6a$, $9ab$, $12b$, $24c$ in that paper are not strongly centred.  They become strongly centred if $\Pi$ is replaced with $D\vpu{\dagger}_{n,2n} \Pi D^{\dagger}_{n,2n}$, where $n = d/3$.  This transformation leaves the stability group unchanged.  Furthermore, if $\Pi$ is strongly-centred then in every known case the set of overlaps $\Tr(D_{\mbf{p}} \Pi$ generates  $\fd{E}_1$ over $\fd{Q}$.
}.  

\vspace{0.1 cm}

We refer to a set of $\EC(d)$ orbits sharing the same SIC field as a multiplet. 

\vspace{0.1 cm}

\noindent \textbf{Observation 9.}    In every known case the Galois group acts transitively on the elements of a multiplet.

\vspace{0.1 cm}

\noindent \textbf{Observation 10.} In those cases where Scott and Grassl~\cite{Scott:2010a} give exact fiducials for what is probably the full set of $\EC(d)$ orbits there are at most two distinct multiplets.   We refer to the field of lowest (respectively highest) degree  as the minimal (respectively maximal) field, and to the corresponding multiplet as the minimal (respectively maximal) multiplet.    It is found that when there are two multiplets, so that the minimal and maximal fields are distinct, the minimal field is a subfield of the maximal field.  

\vspace{0.1 cm}

\noindent \textbf{Observation 11.}
Let $\fd{E}$ be the minimal field in dimension $d$, and  let $\fd{Q}\vartriangleleft \fd{Q}(\sqrt{D}) \vartriangleleft  \fd{E}_0 \vartriangleleft \fd{E}_1 \vartriangleleft \fd{E}$ be the associated tower.  In every known case\cite{Appleby:2016,Appleby:2017a}
\begin{enumerate}
\item $\fd{E}$, $\fd{E}_1$ are ray class fields over $\fd{Q}(\sqrt{D})$ for which the finite part of the  conductor is $d'$,
\item  $\fd{E}$ is the ray class field with ramification allowed at both infinite places,
\item $\fd{E}_1$  is the ray class field with ramification only allowed  at  the infinite  place taking $\sqrt{D}$  to a positive real number,
\item $\fd{E}_0$ is the Hilbert class field over $\fd{Q}(\sqrt{D})$.
\end{enumerate}

\vspace{0.1 cm}

We next describe the action of $\gal(\fd{E}/\fd{Q})$ on the Clifford unitaries.  For any operator $A$ in $\fd{E}$  define 
\ea{
g(A) = \sum_{r,s=0}^{d-1} g\bigl(\la r | A | s\ra\bigr) |r \ra \la s |
} 
where $|0\ra, \dots, |d-1\ra$ is the standard basis. 
Let $k_g$ be the unique integer in the interval $(0,d')$ which is co-prime to $d'$ and such that $g(\tau) = \tau^{k_g}$, and let 
\ea{
H_g &= \bmt 1 & 0 \\ 0 & k_g \emt.
}
Then $g$ acts on the elements of $\C(d)$ according to
\ea{
g(D_{\mbf{p}}U_F) \dot{=} D_{H_g \mbf{p}} U_{H_g F H_g^{-1}}.
}


\vspace{0.1 in}

\noindent \textbf{Observation 12.} Let $\{o_1, \dots, o_l\}$ be the $\EC(d)$ orbits of any of the known multiplets in dimension $d$.  Let $\fd{Q} \vartriangleleft \fd{E}_c \vartriangleleft \fd{E}_0 \vartriangleleft \fd{E}_1 \vartriangleleft \fd{E}$ be the associated tower, and let $\Pi_j$ be a strongly centered fiducial on orbit $o_j$.  Then one finds that for each $g\in \gal(\fd{E}/\fd{E}_c)$ and each index $j$ there exists $F_{h,j} \in \ESL(d)$ and an index $k$ such that
\ea{
g(\Pi\vpu{\dagger}_j ) & = U\vpu{\dagger}_{F_{g,j}} \Pi\vpu{\dagger}_k U^{\dagger}_{F_{g,j}}.
\label{eq:gPiAction}
}
Also
\ea{
g\bigl( \Tr(D_{\mbf{p}} \Pi_j)\bigr) &= \Tr(D_{G_{g,j} \mbf{p}}\Pi_k)
\label{eq:gOlpAction}
}
for all $\mbf{p}$, where 
\ea{
G\vpu{-1}_{g,j} = (\Det F\vpu{-1}_{g,j}) F^{-1}_{g,j} H\vpu{-1}_g.
\label{eq:GjTermsFj}
}  
If $g \in \fd{E}_0$ then $k=j$ in Eqs.~\eqref{eq:gPiAction} and \eqref{eq:gOlpAction}.  
The matrices $F_{h,j}$, $G_{h,j}$ are not unique:  Indeed, they can be replaced with arbitrary elements of the cosets $F_{h,j}S_0(\Pi)$, $S(\Pi)G_{h,j} $ respectively, where
\ea{
S(\Pi) &= \{ (\Det F) F \colon F \in S_0(\Pi)\}.
\label{eq:SPiDef}
}
For the automorphism $\bar{g}_1$ (as defined in Observation 7) we can  choose\cite{Appleby:2013}
\ea{
G_{\bar{g}_1,j} &= I  & F_{\bar{g}_1,j} &= \bmt -1 & 0 \\ 0 & 1 \emt .
\label{eq:fgbarDef}
}
for all $j$.

\vspace{0.1 in}

\noindent Let $\fd{Q} \vartriangleleft \fd{E}_c \vartriangleleft \fd{E}_0 \vartriangleleft \fd{E}_1 \vartriangleleft \fd{E}$ be a tower to which the previous observations apply. With notation as in Observation 12, choose $g\in \gal(\fd{E}_1/\fd{E}_0)$ define $G_{g,j} = G_{g',j}$ where $g'$ is a lift of $g$ to $\gal(\fd{E}/\fd{E}_0)$.  It is easily seen that this definition does not depend on which of the two lifts is taken.  Then it can be shown~\cite{Appleby:2013} that the $G_{g,j}$ are all in $N(\Pi)$, the normalizer of $S(\Pi)$ in $\GL(2,\fd{Z}/d'\fd{Z})$, and that for each fixed $j$ the map $g\to G_{g,j} S(\Pi)$ is an injective homomorphism of $\gal(\fd{E}_1/\fd{E}_0)$ into $N(\Pi)/S(\Pi)$.  It is natural to ask, what is the range of this homomorphism.  The answer depends on whether the $\EC(d)$ orbit is type-$z$ or type-$a$.

\vspace{0.1 in}

\noindent \textbf{Observation 13.}
Suppose the orbit is type-$z$.  Then, in every known case, one finds that the range of the homomorphism is $C(\Pi)/S(\Pi)$, where $C(\Pi)$ is the centralizer of $S(\Pi)$ in $\GL(2,\fd{Z}/d'\fd{Z})$.  So we have an isomorphism
\ea{
\gal(\fd{E}_1/\fd{E}_0) \cong C(\Pi)/S(\Pi).
\label{eq:CSiso}
}

\vspace{0.1 in}

\noindent The conjecture, that this isomorphism is generally true of a type-$z$ orbit, plays an important role in the calculations in the next section.  Its restriction to the case of minimal SIC fields would, if true,  also be of  number-theoretical interest:  For  it would mean that the Galois group of a ray class field over the Hilbert class field of a real quadratic field is, in many cases, isomorphic to the quotient of two matrix groups.

Turning to type-$a$ orbits one has:

\vspace{0.1 in}

\noindent \textbf{Observation 14.}  For the two type-$a$ orbits for which exact solutions are given in ref.~\cite{Scott:2010a} one finds~\cite{Appleby:2013}
\ea{
\gal(\fd{E}_1/\fd{E}_0) \cong C(\Pi)/S(\Pi).
\label{eq:E1ByE0Iso}
}
just as is the case for the type-$z$ orbits.

\vspace{0.1 in}

However this observation is of limited significance.  The problem~\cite{Appleby:2013} is that  one can see from the numerical data that the group $C(\Pi)/S(\Pi)$ is non-Abelian\footnote{By contrast, it can be shown~\cite{Appleby:2013} that for a type-$z$ orbit $C(\Pi)/S(\Pi)$ is  necessarily Abelian.} for  the type-$a$ orbits $21d$, $30d$, $39gh$, $48e$.  So for these orbits one of two things must be true:  Either $\fd{E}$ is not an Abelian extension of $\fd{Q}(\sqrt{D})$, or else $\gal(\fd{E}_1/\fd{E}_0)$ is not isomorphic to $C(\Pi)/S(\Pi)$.  We discuss this problem further in the next section.  


\section{Type-\texorpdfstring{$a$}{a} fiducials}
\label{sec:typea}
An alternative to the isomorphism of Eq.~\eqref{eq:CSiso} holding for type-$a$ fiducials was conjectured in ref.~\cite{Appleby:2013}.  In this section we propose a stronger conjecture, which holds for the previously known fiducials $12b$, $48g$, and also for fiducial $21d$ calculated using the methods described in  Section~\ref{sec:calc}.  


Suppose $d=3 \mod 9$.  Then $d=3n$ where $n=1 \mod 3$ and
\ea{
F_a &= \bmt 1 & 3n + 3 \\ 4n-1& 3n-2\emt.
}
The fact that $n$ is coprime to 3 means (see Appendix B of ref.~\cite{Appleby:2012}) that there is a natural isomorphism
\ea{
\chi \colon \SL(2, \fd{Z}/d'\fd{Z}) \to \SL(2,\fd{Z}/n'\fd{Z}) \times \SL(2,\fd{Z}/3\fd{Z}).  
}
It is straightforward to show that $\chi$ extends to an isomorphism of $\GL(2,\fd{Z}/d'\fd{Z})$ onto $\GL(2,\fd{Z}/n'\fd{Z}) \times \SL(2,\fd{Z}/3\fd{Z})$, and that, for arbitrary $\left( \begin{smallmatrix} \alpha & \beta \\ \gamma & \delta \end{smallmatrix}\right)\in \GL(2,\fd{Z}/d'\fd{Z})$, 
\ea{
\chi \Biggl( \bmt \alpha & \beta \\ \gamma & \delta \emt \Biggr)
=
\Biggl( \bmt \alpha &  3\beta \\  \frac{(2n+1)\gamma}{3} & \delta \emt , \bmt \alpha & \beta \\ \gamma & \delta \emt \Biggr)
}
where $n'=n$ (respectively $n'=2n$) if $n$ is odd (respectively even).  Applying this to $F_a$ we find that 
\ea{
\chi(F_a) &= (\bar{F}_{a}, I)
\label{eq:FaDecomp}
}
where
\ea{
\bar{F}_{a} &= \bmt 1 & n+9 \\ \frac{4n-1}{3} & n-2 \emt .
}
Observe that $\Tr(\bar{F}_{a})= -1 \mod n$, implying that $\bar{F}_{a}$ is conjugate to the Zauner matrix in dimension $n$.

Now consider the isomorphism of  Eq.~\eqref{eq:CSiso}.
Suppose, to begin with, that $\Pi$ is a strongly centred type-$z$ fiducial.  Without loss of generality we may assume that $\Pi$ is an eigenvector of $U_{F_z}$.  Assume also that $S(\Pi)$ is Abelian (as is the case for every known fiducial with $d\ge 4$).  It is shown in ref.~\cite{Appleby:2013} that $C(F_z)$, the centralizer of  $F_z$ in $\GL(2,\fd{Z}/d'\fd{Z})$, is Abelian, from which it follows that $C(\Pi) = C(F_z)$ so that the isomorphism becomes
\ea{
\gal(\fd{E}_1/\fd{E}_0) \cong C(F_z) /S(\Pi).
\label{eq:CSisoB}
}
On the assumption that the left hand side is Abelian (as is the case for every known exact fiducial) this relation cannot hold for an arbitrary type-$a$ orbit.  
Indeed, it can be seen from Eq.~\eqref{eq:FaDecomp} that 
\ea{
C(F_a)\cong C(\bar{F}_{a}) \times \GL(2,\fd{Z}_3).
}
The fact that $\bar{F}_{a}$ is conjugate to the Zauner matrix in dimension $n$ means that $C(\bar{F}_{a})$ is Abelian; however $\GL(2,\fd{Z}_3)$ is non-Abelian, implying that $C(F_a)$ is non-Abelian.  This means that, if $S(\Pi)$ is generated by $F_a$ (as it is in, for example, orbits $21d$, $30d$, $39gh$, $48e$), then $C(F_a)/S(\Pi)$  is non-Abelian.  

As a generalization of Eq.~\eqref{eq:CSisoB} holding for every orbit, irrespective of whether it is type-$z$ or type-$a$, we propose:
\begin{conjecture}\label{cnj:strong} Let $\Pi$ be a strongly-centred fiducial, and let $F\in S(\Pi)$ be such that the unitary $U_F$ is canonical order 3.  Then
\ea{
\gal(\fd{E}_1/\fd{E}_0) \cong \mathcal{M}/S(\Pi).
\label{eq:CSisoC}
}
where $\mathcal{M}$ is a maximal Abelian subgroup of $\GL(2,\fd{Z}/d'\fd{Z})$ containing $F$.
\end{conjecture}
\noindent Note that in the case of type-$z$ fiducials this conjecture reduces to Eq.~\eqref{eq:CSisoB}.  It holds for every known exact fiducial.  In particular, it holds for the type-$a$ fiducials $12b$ and $48g$ (calculated in ref.~\cite{Scott:2010a})  and $21d$ (calculated here). 

In the case of a type-$z$ fiducial there is exactly one maximal Abelian subgroup $\mathcal{M}$ containing $F$; namely, the centralizer $C(F)$.  But for a type-$a$ fiducial there are several.  Indeed, one sees from Eq.~\eqref{eq:FaDecomp} that a subgroup $\mathcal{M}$ containing $F_a$ is maximal Abelian if and only if 
\ea{
\chi(\mathcal{M}) &= C(\bar{F}_{a}) \times \bar{\mathcal{M}}
}
where $\bar{\mathcal{M}}$ is an arbitrary maximal Abelian subgroup of $\GL(2,\fd{Z}_3)$.  One finds that $\bar{\mathcal{M}}$ must be conjugate to one of the three groups
\ea{
\mathcal{\bar{H}}_4 &= \left< \bmt -1 & 0 \\ 0 & -1 \emt, \bmt 1 & 0 \\ 0 & -1 \emt \right> ,
\\
\mathcal{\bar{H}}_6 &= \left< \bmt -1 & 0 \\ 1 &  -1  \emt\right>,
\\
\mathcal{\bar{H}}_8 &= \left< \bmt 1 & -1 \\ 1 & 1 \emt \right>,
}
which are order $4$, $6$ and $8$ respectively.  Consequently, $\mathcal{M}$ must  be conjugate to one of the three groups 
\ea{
\mathcal{H}_j &= \chi^{-1}\bigl(C(\bar{F}_a)\times \bar{\mathcal{H}}_j\bigr), & j&= 4,6,8.
}
We  say that an orbit is type-$a_j$  if $\gal(\fd{E}_1/\fd{E}_0)$ is isomorphic to $\mathcal{H}_j$.  Of the known cases  orbit $12b$ is type-$a_4$ while  orbits $21e$, $48g$ are type-$a_8$.  It is an open question, whether there exist any type-$a_6$ orbits.

In ref.~\cite{Appleby:2012}  the following, weaker conjecture was proposed for strongly-centred type-$a$ fiducials:
\begin{conjecture} \label{cnj:weak}
Let $\Pi$ be a strongly-centred type-$a$ fiducial and let $F\in S$ be such that $U_F$ is a canonical order 3 unitary.  Then 
\ea{
\gal(\fd{E}_1/\fd{E}_0) \cong \mathcal{A}/S(\Pi)
}
where 
\ea{
\mathcal{A} &= \{r I + s G \colon r, s \in \fd{Z}/d'\fd{Z} \text{ and }  rI + s G \in \GL(2,\fd{Z}/d'\fd{Z}\}
}
for some matrix  $G$ such that $F = I + 3 G$.  
\end{conjecture}
\noindent To see that this conjecture  is indeed  weaker, let $\Pi$ be a strongly centred  type-$a$ fiducial satisfying Conjecture~\ref{cnj:strong}. Then  $\gal(\fd{E}_1/\fd{E}_0) \cong \mathcal{H}_j/S(\Pi)$ for some $j$.  The fact that $\bar{F}_a$ is conjugate to the Zauner matrix in dimension $n$ means,  in view of Lemma~12 in ref.~\cite{Appleby:2013},  
\ea{
C(\bar{F}_a) &= \{ r I + s\bar{F}_a \colon r,s \in \fd{Z}/n'\fd{Z} \text{ and } r I + s\bar{F}_a \in \GL(2,\fd{Z}/n'\fd{Z})\}.
}
Define 
\ea{
\bar{G} &= \frac{2n+1}{3}(\bar{F}_a - I)
}
(note that the fact that $n=1 \mod 3$ means $(2n+1)/3$ is an integer).  
Then we also have $\bar{F}_a = I + 3 \bar{G}$, implying
\ea{
C(\bar{F}_a) &= \{ r I + s\bar{G} \colon r,s \in \fd{Z}/n'\fd{Z} \text{ and } r I + s\bar{G} \in \GL(2,\fd{Z}/n'\fd{Z})\}.
}
It is straightforward to show that 
\ea{
\mathcal{\bar{H}}_j &= \{ r I + s\bar{H}_j \colon r,s\in \fd{Z}/3\fd{Z} \text{ and } r I + s\bar{H}_j \in \GL(2,\fd{Z}/3\fd{Z}\}
}
where
\ea{
\bar{H}_4 &= \bmt 1 & 0 \\ 0 & -1\emt,
&
\bar{H}_6 &= \bmt -1 &0 \\ 1 &-1\emt,
&
\bar{H}_8 &= \bmt 1 &-1 \\ 1 & 1 \emt.
}
Using the Chinese Remainder Theorem~\cite{Ireland:1990} one finds
\ea{
\mathcal{H}_j &= \{ rI + s H_j \colon  r,s\in \fd{Z}/d'\fd{Z} \text{ and } r I + s H_j \in \GL(2,\fd{Z}/d'\fd{Z})\}
}
where
\ea{
H_j &= \chi^{-1}(\bar{G},\bar{H}_j).
}
The fact that
\ea{
\chi(I+ 3 H_j) &= (I + 3 \bar{G}, I) = \chi(F_a),
}
means $I + 3 G_j = F_a$.  So $\Pi$ satisfies Conjecture~\ref{cnj:weak}.  To see that Conjecture~\ref{cnj:weak} is strictly weaker than Conjecture~\ref{cnj:strong} consider the order $2$ subgroup of $\GL(2,\fd{Z}/3\fd{Z})$.  
\ea{
\bar{\mathcal{H}}_2 &= \la - I \ra =\{ r I + s \bar{H}_2 \colon r,s\in \fd{Z}/3\fd{Z} \text{ and } r I + s \bar{H}_2 \in \GL(2,\fd{Z}/3\fd{Z})\}
}
where $\bar{H}_2 = - I$.  Let $\mathcal{H}_2 = \chi^{-1}\bigl((\bar{G},\bar{\mathcal{H}}_2)\bigr)$.  Then
\ea{
\mathcal{H}_2 &= \{ rI + sH_2 \colon r,s\in \fd{Z}/d'\fd{Z} \text{ and } rI + sH_2 \in \GL(2,\fd{Z}/d'\fd{Z})\}
\label{eq:mclH2DefA}
}
where
\ea{
 H_2= \chi^{-1} (\bar{G}, \bar{H}_2).
 }
 Moreover $I+3 H_2 =F_a$.  So a fiducial for which $\gal(\fd{E}_1/\fd{E}_0) \cong \mathcal{H}_2/S(\Pi)$ would satisfy Conjecture~\ref{cnj:weak}.  However, it would not satisfy Conjecture~\ref{cnj:strong} because $\mathcal{H}_2$ is not maximal Abelian, being properly contained in $\mathcal{H}_4$,  $\mathcal{H}_6$, $\mathcal{H}_8$.
 
 The group $\mathcal{H}_2$ plays an important role in the reconstruction of exact type-$a$ fiducials from numerical ones, as we will discuss in the next section.  This is because $\bar{\mathcal{H}}_2$ is the centre of $\GL(2,\fd{Z}/3\fd{Z})$, which means that $\mathcal{H}_2$ is contained in every maximal Abelian subgroup of $\GL(2,\fd{Z}/d'\fd{Z})$ containing $F_a$.  In view of its importance it may be worth noting that, in addition to Eq.~\eqref{eq:mclH2DefA}, one also has
 \ea{
\mathcal{H}_2 &= \{ rI + sF_a \colon r,s\in \fd{Z}/d'\fd{Z} \text{ and } rI + sF_a\in \GL(2,\fd{Z}/d'\fd{Z})\}.
\label{eq:mclH2DefB}
 }
 Indeed, it is easily verified that 
 \ea{
 H_2 &= \frac{2n+1}{3} F_a + \frac{4n-1}{3} I
 }
 (note that the fact that $n=1 \mod 3$ means $2n+1$ and $4n-1$ are both divisible by $3$).  Together with the relation $F_a = I + 3 H_2$ this means
\ea{
&\{ r I + s H_2 \colon r, s\in \fd{Z}/d'\fd{Z} \text{ and } r I + s H_2 \in \GL(2,\fd{Z}/d'\fd{Z})\} 
\nn
&\hspace{0.5 in} = \{ r I + s F_a \colon r, s\in \fd{Z}/d'\fd{Z} \text{ and } r I + s F_a \in \GL(2,\fd{Z}/d'\fd{Z})\}.
}
 
 \section{Fields and Multiplets}
 \label{sec:FieldsMultiplets}
 It is shown in ref.~\cite{Appleby:2016} that in every known case the minimal SIC field in each dimension is the ray-class field over $\fd{Q}(\sqrt{D})$ with conductor $d'$ and ramification at both infinite places.  We will refer to the conjecture, that this is always the case, as the ray-class conjecture. However in many dimensions (not all) there are other SIC fields, having higher degree.  This raises the question, whether one can generalize the ray-class conjecture to these additional fields.  
 
 For the dimensions where Scott and Grassl give a full set of exact solutions there are either one or two SIC fields.  We refer to the field of lowest (respectively highest) degree as the minimal field, denoted $\fd{E}_{\rm{min}}$ (respectively maximal field, denoted $\fd{E}_{\rm{max}}$).  Associated to the two fields are two multiplets, which we refer to as the minimal and maximal multiplets.  This information is presented in Table~\ref{tbl:multipletSG}. 
\begin{table}[htb]
\begin{center}
\begin{tabular}{cccc}
\toprule
$d$ & minimal multiplet &  maximal multiplet & $[\fd{E}_{\rm{max}} \colon \fd{E}_{\rm{min}}]$
\\
\midrule
$4$ & $4a$ && 
\\
$5$ & $5a$ &&  
\\
$6$ & $6a$ &&  
\\
$7$ & $7b$  &  $7a$ & $2$
\\
$8$ & $8b$ &  $8a$ &  $4$
\\
$9$ & $9ab$  &  & 
\\
$10$ & $10a$  &&
\\
$11$ & $11c$  & $11ab$ & $2$
\\
$12$ & $12b$ & $12a$ & $3$
\\
$13$& $13ab$ & & 
\\
$14$ & $14ab$ &&
\\
$16$ & $16ab$ && 
\\
\bottomrule
\end{tabular}
\end{center}
\caption{\label{tbl:multipletSG}  Minimal and maximal multiplets for dimensions where Scott and Grassl~\cite{Scott:2010a} give a full set of exact fiducials.  We only list the maximal multiplet when it is distinct from the minimal one.  The right-most column gives the degree of the extension $\fd{E}_{\rm{max}}/\fd{E}_{\rm{min}}$, where $\fd{E}_{\rm{max}}$ is the maximal field and $\fd{E}_{\rm{min}}$ is the minimal one.}
\end{table}
It raises several questions:
\begin{enumerate}
\item In the four dimensions listed for which there are two distinct SIC fields, the maximal field is an extension of the minimal one.  One would like to know if that is merely a low dimensional accident, or whether it is generally true.  
\item There are at most two distinct fields in the dimensions listed.  One would like to know if in higher dimensions there are sometimes more than two.
\item In the dimensions listed it never happens that there is more than one multiplet associated to a given field.  One would like to know if this $1:1$ correspondence between fields and multiplets persists in higher dimensions.
\item In the dimensions listed $\fd{E}_{\rm{min}}$  is the ray class field over $\fd{Q}(\sqrt{D})$ with conductor $d'$ and ramification at both infinite place.  One would like to formulate a  conjecture applying to the non-minimal fields in each dimension.  For this purpose it would be useful to have more examples than the four cases $7a$, $8a$, $11ab$, $12a$.
\end{enumerate}
One of our aims in the calculations reported here was to address these questions.  Our results reveal that the number of fields in a given dimension can be larger than $2$.  However, in every case calculated, one continues to find that there is a unique  field $\fd{E}_{\rm{min}}$ of minimal degree over $\fd{Q}$, and a unique  field $\fd{E}_{\rm{max}}$ of maximal degree over $\fd{Q}$.  The field $\fd{E}_{\rm{min}}$ is always the ray class field over $\fd{Q}(\sqrt{D})$ with conductor $d'$ and ramification at both infinite places.  The  field $\fd{E}_{\rm{max}}$ always contains $\fd{E}_{\rm{min}}$, and additional fields $\fd{E}$, when they exist, always lie between these two:
\ea{
\fd{E}_{\rm{min}} \subseteq \fd{E} \subseteq {E}_{\rm{max}}.
}
We accordingly refer to them as intermediate fields, and to the associated multiplets as intermediate multiplets.  Finally, one continues to find that there is a $1:1$ correspondence between fields and multiplets.  This information is summarized  in Table~\ref{tbl:multiplet} and Fig.~\ref{fg:multiplet}
\begin{table}[htb]
\begin{center}
\begin{tabular}{ccccc}
\toprule
$d$ & \parbox[c][][c]{1.5 cm}{minimal multiplet} & \parbox[c][][c]{2 cm}{intermediate multiplet(s)} & \parbox[c][][c]{1.5 cm}{maximal multiplet} & $[\fd{E}_{\rm{max}} \colon \fd{E}_{\rm{min}}]$
\\
\midrule
$15$ &  $15d$ &  $15b$ & $15ac$ & $4$
\\
$17$ &  $17c$ &  & $17ab$ & $2$
\\
$18$ & $18ab$ & & 
\\
$19$ & $19e$ & $19a$, $19d$ & $19bc$ & $12$
\\
$20$ & $20ab$ &&
\\
$21$ & $21e$ && $21abcd$ & $3$
\\
$24$ & $24c$  && $24ab$  & $4$
\\
$28$ &  $28c$ && $28ab$  & $4$
\\
$30$ & $30d$ &  & $30abc$  & $3$
\\
$35$ & $35j$ & $35i$, $35e$,  $35h$, $35af$ & $35bcdg$ & $16$
\\
$39$ & $39ij$ & $39bf$, $39gh$ & $39acde$ & $6$
\\
$48$ & $48g$ & $48e$, $48f$ &  $48abcd$ & $24$
\\
\bottomrule
\end{tabular}
\end{center}
\caption{\label{tbl:multiplet} Minimal, maximal and intermediate multiplets for the  dimensions calculated in this paper.  Note that exact fiducials for orbits $15d$, $19e$, $24c$, $28c$, $35j$, $48g$ were already known;  all other exact fiducials in these dimensions are new, however.  }
\end{table}
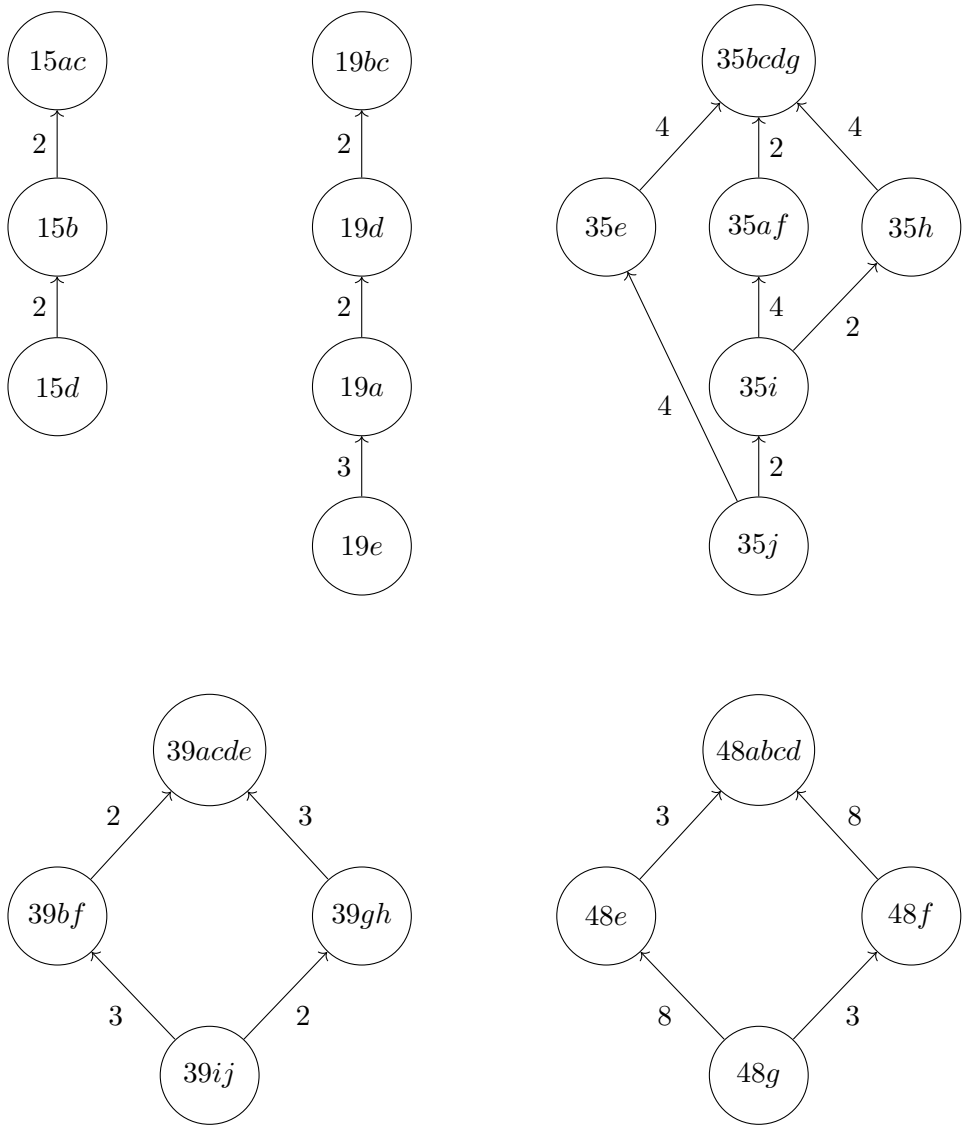
\begin{figure}[htb]
\begin{tikzpicture}
[auto, nd/.style={circle,minimum size = 13 mm,draw}]

\matrix[row sep=8mm,column sep=6mm] {
\node (15ac) [nd] {$15ac$};&&\node (19bc) [nd] {$19bc$}; &[7 mm] & &\node (35bcdg) [nd] {$35bcdg$};
\\
\node (15b) [nd] {$15b$}; &&\node (19d) [nd] {$19d$} ; &&\node  (35e) [nd] {$35e$}; & \node (35af) [nd] {$35af$}; & \node (35h) [nd] {$35h$};
\\
\node(15d)  [nd] {$15d$}; &&\node (19a) [nd] {$19a$}; & &&\node (35i) [nd] {$35i$};
\\
&& \node (19e) [nd] {$19e$}; & &&\node (35j) [nd] {$35j$};
\\[5 mm]
& \node (39acde) [nd] {$39acde$}; & & & & \node (48abcd) [nd] {$48abcd$};
\\
\node (39bf) [nd] {$39bf$};  & & \node (39gh) [nd] {$39gh$}; & & \node (48e) [nd] {$48e$}; & & \node (48f) [nd] {$48f$};
\\
& \node (39ij) [nd] {$39ij$}; &&&& \node (48g) [nd] {$48g$};
\\
};
\draw[->] (15d) to node{2} (15b);
\draw[->] (15b) to node{2} (15ac);
\draw[->] (19e) to node{3} (19a);
\draw[->] (19a) to node{2} (19d);
\draw[->] (19d) to node{2} (19bc);
\draw[->] (35j) to node[swap] {2} (35i);
\draw[->] (35i) to node[swap] {2} (35h);
\draw[->] (35j) to node{4} (35e);
\draw[->] (35i) to node[swap] {4} (35af);
\draw[->] (35e) to node{4} (35bcdg);
\draw[->] (35af) to node[swap] {2} (35bcdg);
\draw[->] (35h) to node[swap] {4} (35bcdg);
\draw[->] (39ij) to node{3} (39bf);
\draw[->] (39ij) to node[swap] {2} (39gh);
\draw[->] (39bf) to node{2} (39acde);
\draw[->] (39gh) to node[swap] {3} (39acde);
\draw[->] (48g) to node {8} (48e);
\draw[->] (48g) to node[swap] {3} (48f);
\draw[->] (48e) to node {3} (48abcd);
\draw[->] (48f) to node[swap] {8} (48abcd);
\end{tikzpicture}
\caption{\label{fg:multiplet}Field inclusions for cases where there are one or more intermediate fields.  The arrows run from the smaller field to the larger; numbers beside the arrows are the degrees of the extensions.}
\end{figure}
\section{Method for Calculating  Exact Fiducials}
\label{sec:calc}
We first describe the obvious, or ``brute-force'' approach to the problem.  We then describe the refinements  actually used to perform the calculations in this paper.  
Suppose that $\Pi_N$ is a known high precision numerical fiducial corresponding to an unknown exact fiducial $\Pi_E$ in dimension $d$.  Suppose, in addition, that it is somehow known that $\fd{Q}(\Pi_E,\tau)$ is the minimal SIC-field for dimension $d$.  If the conjectures in ref.~\cite{Appleby:2016,Appleby:2017a} are correct $\fd{Q}(\Pi_E,\tau)$ is then the ray class field over $\fd{Q}(\sqrt{D})$ with conductor $d'$ and ramification at both infinite places.  Using \emph{Magma} we can easily calculate a basis for the field over $\fd{Q}$.  Let $b_1, \dots, b_n$ be such a basis.   Then for all $j$, $k = 0 , 1 , \dots , d-1$ there exist integers $\mu_0(j,k), \mu_1(j,k), \dots, \mu_n(j,k)$ such that
\ea{
\mu_0(j,k) \la j |\Pi_E |k \ra -\sum_{r=1}^n \mu_r(j,k) b_r = 0.
}
 For the numerical fiducial one then has 
\ea{
\mu_0(j,k) \la j |\Pi_N |k \ra -\sum_{r=1}^n \mu_r(j,k) b_r \approx 0.
}
to a high degree of precision.  If the precision is sufficiently great we can use an integer relation algorithm to find the integers $\mu_r(j,k)j$.  Substituting these integers in
\ea{
\sum_{r=1}^n \frac{\mu_r(j,k)}{\mu_0(j,k)} b_r
}
 gives us the exact matrix element $ \la j |\Pi_E |k \ra$ for each $j$, $k$.  There are three problems with this ``brute-force'' approach:
\begin{itemize}
\item It requires us to guess which numerical fiducials correspond to an  exact fiducial generating the ray-class field.
\item We do not currently   have a conjecture for  $\fd{Q}(\Pi_E,\tau)$ when it is not the minimal field.
\item In the cases of interest to us the degree of the minimal field is  large, which means that the required precision is large, and the calculation correspondingly slow.
\end{itemize}
To  deal with these problems we therefore modified the procedure.  We actually employed two different methods, which we refer to as methods 1 and 2.  Method 1 was the original method used to calculate the exact fiducials in dimensions $17$--$21$.  Method 2 is a much improved method used to calculate the exact fiducials for $d\ge 24$ (and also for $d=15$ which, although the dimension is smallest, was actually calculated last).  

\subsection*{Method 1}  
We describe the method as it  applies for a type-$z$ fiducial, and then indicate the modifications needed to deal with a type-$a$ fiducial at the end. 
Define
\ea{
\chi_{E,\mbf{p}} &= \Tr\bigl( D_{\mbf{p}} \Pi_E \bigr) & \chi_{N,\mbf{p}} &= \Tr\bigl( D_{\mbf{p}} \Pi_N \bigr)
}
Let $C(\Pi_E)$ be the centralizer of the symmetry group $S(\Pi_E)$, and let $\mcl{O}_1\dots \mcl{O}_m$ be the orbits of $\bigl(\fd{Z}/d'\fd{Z}\bigr)^2$ under the action of $C(\Pi_E)$.  If $d\neq 0 \mod 3$ (respectively\footnote{We need to treat the case when $d=0\mod 3$ differently because one cannot tell if a numerical fiducial is strongly centred or not.  If the fiducial is not strongly centred then $g(\chi_{\mbf{p}})$ only equals $\chi_{G_g \mbf{p}}$ up to a cube root of unity~\cite{Appleby:2013} (where $g$ is any element of $\gal(\fd{E}_1/\fd{E}_0)$), and the coefficients of $Q_j(x)$ are consequently guaranteed to be in the field $\fd{E}_0$.  Cubing the overlaps obviates this difficulty.} $d=0\mod 3$) let $\mcl{S}_{E,j}$ be the set of distinct numbers $\chi_{E,\mbf{p}}$ (respectively $\chi^3_{E,\mbf{p}}$)  obtained as $\mbf{p}$ runs over $\mcl{O}_j$, and define
\ea{
Q_{E,j}(x) &= \prod_{s \in \mcl{S}_{E,j}} (x-s) .
\label{eq:QjPolyDef}
}
Of course, we do not know the overlaps $\chi_{E,\mbf{p}}$, so the polynomial $Q_{E,j}(x)$ is also unknown.  However,  $S(\Pi_E)$,  $C(\Pi_E)$  and the orbits $\mcl{O}_1\dots \mcl{O}_m$ can all be calculated just from a knowledge of $\Pi_N$.  Let $\mcl{S}_{N,j}$ be the set of distinct numbers $\chi_{N,\mbf{p}}$ ($\chi^3_{E,\mbf{p}}$ if $d=0\mod 3$)  obtained as $\mbf{p}$ runs over $\mcl{O}_j$, and define
\ea{
Q_{N,j}(x) &= \prod_{s \in \mcl{S}_{N,j}} (x-s) .
\label{eq:QNjPolyDef}
}
The known polynomials $Q_{N,j}(x)$ will then be high precision  approximations to the unknown polynomials $Q_{E,j}(x)$. 
 If the conjecture that $\gal(\fd{E}_1/\fd{E}_0)$ is isomorphic to $C\Pi)/S(\Pi)$, is correct (\emph{c.f.} Eq.~\eqref{eq:CSiso}) then the coefficients of $Q_{E,j}(x)$ all belong to the field $\fd{E}_0$.  Since the degree of  $\fd{E}_0/\fd{Q}$ is much less than that of  $\fd{E}/\fd{Q}$ (no greater than $8$ for the fiducials calculated in this paper), the degree of $Q_{E,j}(x)$, and consequently the degree of its numerical approximant $Q_{N,j}(x)$, is also comparatively small.    It is consequently possible to use an integer relation algorithm to calculate the coefficients of $Q_{E,j}(x)$ from those of $Q_{N,j}(x)$ without knowing the field in advance, and without an impractically high degree of precision and CPU time (using, for instance, the \emph{Magma} function \verb|MinimalPolynomial| or the \emph{Mathematica} function \verb|RootApproximant|).  In practice it is most efficient to use this method to find  the next-to-leading coefficient of the lowest degree polynomial $Q_{E,j} (x)$ (because  this is the coefficient which involves the smallest integers, and is therefore easiest to find).  Once $\fd{E}_0$ has been ascertained one  can then use a function such as \verb|IntegerRelation| (in \emph{Magma}) or \verb|FindIntegerNullVector| (in \emph{Mathematica}) which assumes a knowledge  of the field to find every coefficient of every polynomial.  The total CPU time needed to obtain exact expressions for the full set of polynomials $Q_{E,j}(x)$ was  $\sim 1$ second  or less in every case.  The precision needed was $~10^3$ decimal digits or less.  We then used \emph{Magma} to construct the  the ray class field over $\fd{Q}(\sqrt{D})$ with conductor $d'$ and ramification at both infinite places.  For a minimal SIC fiducial  exact values of the overlaps were found directly, by factoring the $Q_{E,j}(x)$ over this field.  If the fiducial was not minimal we factored the $Q_{E,j}(x)$ as far as possible, and then used \emph{Magma} to find simplified generators for $\fd{E}$ as an extension of the ray class field.  The overlaps can then be expressed in terms of the generators.  This part of the calculation took $\sim1$ day of CPU time.  Finally, the exact fiducial was calculated from the overlaps using Eq.~\eqref{eq:DpBsis}.
 
For a type-$a$ fiducial the above procedure needs to be modified slightly.  This is because, if Conjecture~\ref{cnj:strong} is correct,  $\gal(\fd{E}_1/\fd{E}_0)$ is isomorphic to $\mathcal{M}/S(\Pi_E)$  for some maximal Abelian subgroup $\mathcal{M}$ which, unlike $C(\Pi_E)$ for a type-$z$ fiducial, cannot be determined from the numerical data.  To get round this problem we use the fact that, as shown in Section~\ref{sec:typea}, the group $\mathcal{M}$  necessarily contains the group $\mathcal{H}_2$, which can be calculated in advance of knowing the exact fiducial from Eq.~\eqref{eq:mclH2DefB}.  The calculation then mirrors the calculation for a type-$z$ fiducial, but using $\mathcal{H}_2$ instead of $C(\Pi_E)$.  Specifically,  let $\mcl{O}_1\dots \mcl{O}_m$ be the orbits of $\bigl(\fd{Z}/d'\fd{Z}\bigr)^2$ under the action of $\mathcal{H}_2$, and  let $\mcl{S}_{E,j}$ be the set of distinct numbers $\chi^3_{E,\mbf{p}}$  obtained as $\mbf{p}$ runs over $\mcl{O}_j$;  then define
$Q_{E,j}(x) = \prod_{s\in \mcl{S}_{E,j}} (x-s)$.  Let 
\ea{
f\colon \gal(\fd{E}_1/\fd{E}_0) \to \mathcal{M}/S(\Pi_E)
}
Then the coefficients of $Q_{E,j}(x)$ are  in $\fd{F}$, the fixed field of $f^{-1} (\mathcal{H}_2)$.  It follows from the results in Section~\ref{sec:typea} that $[\mathcal{M} \colon \mathcal{H}_2] \le 4$.  Consequently,  the degree of $\fd{F}/\fd{Q}$ is no more than $4$ times greater than the degree of $\fd{E}_0/\fd{Q}$, which means one can apply the technique described in the last paragraph for a type-$z$ fiducial to determine the coefficients of the $Q_{E,j}(x)$ from their approximants $Q_{N,j}(x)$, and hence to determine the  exact fiducial.  

There are two main steps to the method just described.  The first step is to find the coefficients of the polynomials $Q_{E,j}(x)$ using an integer relation algorithm; the second  is to factor the polynomials.  As we noted above the first step is fast, taking $\lesssim 1$ second  of CPU time in every case tried.  The second step is much slower. Moreover, the computation time  grows rapidly with the degree of the  number field.  To deal with this problem we developed a new method, which does not rely so heavily on factoring, and is therefore much more efficient.  While developing this method we also discovered the fact noted in Section~\ref{sec:prelim}, that every known centred fiducial becomes strongly centred when multiplied by the appropriate displacement operator.  This obviated the need when $d=0\mod 3$ to first calculate the cubed overlaps  and then to take the cube root at the end of the calculation.

\subsection*{Method 2} As before we first describe  the method as it applies to type-$z$ fiducials, the modifications needed to handle type-$a$ fiducials being described afterwards.

As with method 1, we begin by calculating high precision numerical approximations to the polynomials $Q_{E,j}(x)$; with, however, the difference that even when $d=0 \mod 3$ the numbers $s$ in Eq.~\eqref{eq:QNjPolyDef} are  the overlaps themselves, as opposed to their cubes.  We then use the \emph{Magma} function \verb|MinimalPolynomial| to obtain exact versions of the coefficients.  If $d\neq 0 \mod 3$ we take $\fd{E}_0$ to be the field generated by the coefficients.  If $d=0 \mod 3$ we repeat the calculation for each of the nine fiducial vectors $D_p|\psi\ra$ for which $p = 0 \mod d/3$, and then take the desired strongly centred fiducial to be the one for which the polynomials obtained by applying \verb|MinimalPolynomial| to the coefficients have lowest degree.  If the SIC is minimal $\fd{E}_1$ can be taken to be the appropriate ray-class field.  Otherwise $\fd{E}_1$ is an extension of this field.  We find a set of generators for the extension by factoring some of the polynomials $Q_{E,j}(x)$.  It is crucial to the success of this method that in practice one does not need to factor all the $Q_{E,j}(x)$.  More than that:  it turns out that one can get the additional field generators by factoring a polynomial whose degree is order $10$ or less (which particular polynomial being ascertained by trial and error).  The fact that the degree is so small means that the factoring can be done in minutes at most.  If instead one had to factor the $Q_{E,j}(x)$ of highest degree this method would have few, if any advantages over Method 1 described above.

Once the fields $\fd{E}_0$ and $\fd{E}_1$ have been determined we use the conjectured isomorphism of Eq.~\eqref{eq:CSiso} to reduce the rest of the calculation to a problem in linear algebra.  Let $b_1,  \dots, b_n$ be a basis for the extension $\fd{E}_1/\fd{E}_0$,  let $g_1, \dots, g_n$ and $G_1 S(\Pi_E), \dots , G_n S(\Pi_E)$  be  explicit listings of the elements of $\gal(\fd{E}_1/\fd{E}_0)$ and $C(\Pi_E)/S(\Pi_E)$ respectively, and let $\mcl{P}$ be the set of permutations $f$ with the property that
\ea{
g_j \to G_{f(j)} S(\Pi_E)
}
is an isomorphism  of $\gal(\fd{E}_1/\fd{E}_0)$ onto $C(\Pi_E)/S(\Pi_E)$. Given any overlap $\Tr(D_{\mbf{p}} \Pi_E)$
\ea{
  \sum_{k=1}^n s_k b_k&= \Tr(D_{\mbf{p}} \Pi_E)
}
for some $s_k \in \fd{E}_0$.  The conjecture is that for some permutation $f \in \mcl{P}$
\ea{
g_j \bigl( \Tr(D_{\mbf{p}} \Pi_E) \bigr) &= \Tr(D_{G_{f(j)}\mbf{p}} \Pi_E)
}
for all $j$.  If correct this means 
\ea{
S &= B^{-1}V_{f}
\label{eq:Svc}
\\
\intertext{where $B$ is the matrix}
B&=\bmt g_1(b_1) & g_1(b_2) & \dots \\ g_2(b_1) & g_2(b_2) & \dots \\ \vdots & \vdots & \emt
}
and $S$, $V_{f}$ are the vectors
\ea{
S &= \bmt s_1 \\ s_2 \\ \vdots\emt, & V_{f_0} &= \bmt \Tr(D_{G_{f(1)}\mbf{p}} \Pi_E) \\  \Tr(D_{G_{f(2)}\mbf{p}} \Pi_E)  \\ \vdots \emt.
}
Of course, we do not know the permutation $f$.  To find it we consider in turn each of the permutations in $\mcl{P}$, and use it in conjunction with Eq.~\eqref{eq:Svc} to calculate a high-precision candidate for the vector $S$ (in practice $10^3$ digits of precision was sufficient).  Let $s_j$ be the $j^{\rm{th}}$ component of this candidate vector, and let $e_1, \dots, e_m$ be a basis for $\fd{E}_0/\fd{Q}$.  We use the \emph{Magma} function \verb|IntegerRelation|  to find integers $l_{j,0}, \dots, l_{j,m}$  solving the equation $l_{j,0} s_j - l_{j,1} e_1 - \dots - l_{j,m} e_m =0$.  We find that for one choice of permutation $f$ the norms $\sqrt{l^2_{j,0}+\dots + l^2_{j,m}}$ are many orders of magnitude smaller than they are for any of the others.  We take this permutation to correspond to the true isomorphism between $\gal(\fd{E}_1/\fd{E}_0)$ and $C(\Pi_E)/S(\Pi_E)$, and 
\ea{
S =\sum_{a=1}^{m}  e_a \bmt \frac{l_{1,a}}{l_{1,0}} \\ \frac{l_{2,a}}{l_{2,0}} \\  \vdots \emt
}
to be the exact value of the vector $S$.  We then invert Eq.~\eqref{eq:Svc} to obtain the exact value of the overlaps $\Tr(D_{G_{f(j)}\mbf{p}} \Pi_E)$.  By repeating this procedure we obtain an exact value for every overlap and then use Eq.~\eqref{eq:DpBsis} to obtain the exact fiducial.  Finally, we check that the vector so obtained really is an exact fiducial.

In this description we have assumed that the fiducial to be calculated is type-$z$.  To calculate a type-$a$ fiducial the method can be modified along the lines indicated at the end of our description of method 1. 

\appendix
\section{Results}
\label{sec:results}

We used the method described in Section exact fiducials for $\EC(d)$ orbits $17abc$, $18ab$, $19abcd$, $20ab$, $21abcde$.  The fiducials themselves are available online~\cite{Exact:2016}.  The purpose of this appendix is to describe the associated fields, Galois groups and the isomorphism of Eq.~\eqref{eq:CSisoC}.

When $d$ is not a multiple of $3$ we always calculate an exact version of the corresponding Scott-Grassl~\cite{Scott:2010a} numerical fiducial.  When $d$ is a multiple of $3$ we transform the Scott-Grassl fiducial so as to make it strongly-centred.  Let $\Pi_{\rm{sg}}$ be an exact version of  the Scott-Grassl  numerical fiducial in such a dimension, and let $\Pi_{\rm{sc}}$ be the strongly-centred fiducial we have calculated.  Then
\ea{
\Pi_{\rm{sc}} = U \Pi_{\rm{sg}} U^{\dagger}
}
where $U$ is the displacement operator specified in Table~\ref{tbl:sc}.  
\begin{table}[hbt]
\begin{center}
\begin{tabular}{cccccccccc}
\toprule
orbit & $U$ & orbit & $U$ & orbit & $U$ & orbit & $U$ &
\\
\midrule
$15a$ & $D_{5,10}$ & $21bcd$ &  $D_{14,7}$ & $30d$ & $D_{20,20}$ & $39h$ & $D_{26,0}$ 
\\
$15c$ & $D_{10,5}$ & $21e$ & $D_{7,0}$ & $39abcd$ & $D_{26,13}$ & $39i$ & $D_{26,26}$
\\
$15bd$ & $I$  & $24abc$ & $D_{8,16}$ & $39e$ & $I$ & $39j$ & $D_{13,13}$
\\
$18ab$  & $D_{12,6}$ & $30ac$ & $D_{30,10}$ & $39f$ & $D_{13,26}$ & $48adfg$ & $I$
\\
$21a$ &  $D_{7,14}$ & $30b$ & $I$ & $39g$ & $D_{0,13}$ & $48bc$ & $D_{16,32}$
\\
&& && &&  $48e$ & $D_{16,16}$
\\
\bottomrule
\end{tabular}
\end{center}
\caption{\label{tbl:sc} Displacement operators converting Scott-Grassl numerical fiducials to strongly-centred ones in dimensions divisible by $3$.}
\end{table}

The multiplet structure for these dimensions is specified in Table~\ref{tbl:multiplet}.
The maximal field for each dimension is specified in Tables~\ref{tble:artGenerators}--\ref{tbl:minpolys}, using the same notational conventions as in ref.~\cite{Appleby:2013} to denote the field generators:  
\begin{itemize}
\item $a$  denotes $D$, the square-free part of $(d-3)(d+1)$.
\item $r_1, \dots, r_j$  denote square roots of integers.
\item $t$  denotes $\cos \pi/d$ or $\sin \pi/d$.
\item $b_1, \dots, b_k$ denote numbers constructed recursively from $\fd{Q}(a, r_1, \dots, r_j , t)$ by taking sums, products and roots.
\item $i$ denotes the square root of $-1$.
\end{itemize}
The $a$, $r$ and $t$ generators are tabulated in Table~\ref{tble:artGenerators}; the $b$ generators in Tables~\ref{tbl:bgens1}, \ref{tbl:bgens2}, \ref{tbl:bgens3}.  The minimal polynomials of the  cubic $b$ generators are  tabulated in Table~\ref{tbl:minpolys}. The fields in the tower of Eq.~\eqref{eq:tower} are tabulated in Table~\ref{tbl:fields}, for both the maximal and the minimal and intermediate multiplets in each dimension.

Information regarding the Galois group is tabulated dimension by dimension, in a series of boxes.  Let  $\fd{Q} \vartriangleleft \fd{E}_c \vartriangleleft \fd{E}_0 \vartriangleleft \fd{E}_1 \vartriangleleft \fd{E}$ be the tower for the maximal multiplet in dimension $d$.  The first box  specifies the action of the generators of $\gal(\fd{E}/\fd{Q})$.  The generators are denoted $g_a$, $g_1, \dots, g_m$, $\bar{g}_1$ and are chosen so that
\begin{enumerate}
\item $g_a$ is  an order $2$ extension of the non-trivial element of $\gal(\fd{E}_c/\fd{Q})$.
\item $g_1, \dots , g_m , \bar{g}_1$ fix $\fd{E}_c$.  In particular, they are mutually commuting.  Moreover
\ea{
\gal(\fd{E}/\fd{E}_c) &= \la g_1 \ra \oplus \dots \oplus \la g_m \ra \oplus \la \bar{g}_1 \ra
\label{eq:fdEfdEcDsm}
}
\item $\bar{g}_1$ is the non-trivial element of $\gal(\fd{E}/\fd{E}_1)$.
\item Complex conjugation is given by $g_c = g_a \bar{g}_1 g_a$ (\emph{c.f.}\ Eq.~\eqref{eq:gcgb1}).
\end{enumerate}
For each generator $h$, Box $1$ also tabulates the order of $h$, and $g_a h g_a$.  In view of Eq.~\eqref{eq:fdEfdEcDsm} it therefore completely specifies $\gal(\fd{E}/\fd{Q})$.  Box $2$ specifies $\gal(\fd{E}/\fd{Q})$ and its various subgroups for every multiplet in dimension $d$.  It also specifies how the Galois group switches between the different $\EC(d)$ orbits in a multiplet.  Finally, we tabulate the $G$ matrices defined by Eq.~\eqref{eq:gOlpAction}.  In dimension $d$ let  $dx$, $dy$ be two $\EC(d)$ orbits such that generator $g_j$ takes $dx$ to $dy$, and let $\Pi_x$, $\Pi_y$ be the exact fiducials on these orbits given in ref.~\cite{Exact:2016}.  Then
\ea{
g_j \bigl( \Tr( D_{\mbf{p}} \Pi_x)\bigr) &= \Tr(D_{G_{xj} \mbf{p}} \Pi_y)
}
where $G_{xj}$ are the matrices tabulated in box 3.  We only tabulate the matrices for cases where the action is non-trivial, so that $g_j(\Pi_x) \neq \Pi_x$.  The $F$ matrices in Eq.~\eqref{eq:gPiAction} can then be obtained by inverting Eq.~\eqref{eq:GjTermsFj}.

\begin{landscape}

\begin{table}[htb]
\begin{center}
\begin{tabular}{ccccccccc}
\toprule 
$d$ & $a$ & $r_1$ & $r_2$ & $r_3$ & $r_4$ & $ir_3$ & $t$ & minimal polynomial of $t$ over $\fd{Q}(a,r_1, \dots, r_j)$
\\
\midrule  \addlinespace[4 pt]
$15$ & $\sqrt{3}$ & $\sqrt{5}$  &  & &  & & $\cos\frac{\pi}{15}$ & $8x^2-2(r_1-1)x-(r_1+3)$
\\ \addlinespace[4 pt]
$17$ & $\sqrt{7}$ & $\sqrt{3}$ & $\sqrt{17}$ & &&& $\sin \frac{\pi}{17}$ & $x^8 +\frac{1}{8} (r_2 - 17)x^6 -\frac{1}{32} (7r_2 -51)x^4 +\frac{1}{64} (7r_2 - 34)x^2 -\frac{1}{256}(4r_2 -17)$
\\ \addlinespace[4 pt]
$18$ & $\sqrt{285}$ & $\sqrt{5}$ & $\sqrt{3}$ & & && $\cos\frac{\pi}{18}$ & $x^3 - \frac{3}{4} x - \frac{1}{8} r_2$
\\ \addlinespace[4 pt]
$19$ & $\sqrt{5}$ & $\sqrt{19}$ & $\sqrt{2}$ & && & $\cos\frac{\pi}{19}$ & $512x^9 - 256x^8 - 1024x^7 + 448x^6 + 672x^5 - 240x^4 - 160x^3 + 40x^2 + 10x - 1$
\\ \addlinespace[4 pt]
$20$ & $\sqrt{357}$ & $\sqrt{17}$ & $\sqrt{3}$ & $\sqrt{6}$ & $\sqrt{5}$ & & $\cos\frac{\pi}{20}$ &
$x^2 - \frac{1}{12}r_2r_3(r_4 +1)x + \frac{1}{8}(r_4 - 1)$
\\ \addlinespace[4 pt]
$21$ & $\sqrt{11}$ & $\sqrt{3}$ & $\sqrt{7}$ & & && $\cos\frac{\pi}{21}$ & $x^3-\frac{1}{4}(r_1r_2-1)x^2-\frac{1}{8}(r_1r_2+1)x+\frac{1}{16}(r_1r_2+5)$
\\ \addlinespace[4 pt]
$24$ & $\sqrt{21}$ & $\sqrt{2}$ & $\sqrt{3}$ & $\sqrt{5}$ & &&  $\cos\frac{\pi}{24}$ & $8x^2-(4+r_1+r_1 r_2)$
\\ \addlinespace[4 pt]
$28$ & $\sqrt{29}$ & $\sqrt{2}$ & $\sqrt{7}$ & $\sqrt{5}$ && & $\cos\frac{\pi}{28}$ & $16x^3-4r_1(r_2-1)x^2-4(r_2+1)x+r_1(r_2+3)$
\\ \addlinespace[4 pt]
$30$ & $\sqrt{93}$ & $\sqrt{3}$ & $\sqrt{5}$ & & && $\cos\frac{\pi}{30}$ & $8x^2 -2r_1 (r_2 +1)x + r_2 + 1$
\\ \addlinespace[4 pt]
$35$ & $\sqrt{2}$ & $\sqrt{5}$ & $\sqrt{7}$ & $\sqrt{3}$ & && & 
\\ \addlinespace[4 pt]
$39$ & $\sqrt{10}$ & $\sqrt{20}$ & $\sqrt{13}$ & & & $i\sqrt{3}$ & &
\\ \addlinespace[4 pt]
$48$ & $\sqrt{5}$ & $\sqrt{2}$ & $\sqrt{3}$ & $\sqrt{105}$ & && $\cos\frac{\pi}{48}$ & $32x^4-32x^2+(4-r_1-r_1r_2)$
\\
\bottomrule
\end{tabular}
\end{center}
\caption{$a$, $r$ and $t$ generators\label{tble:artGenerators}}
\end{table}

\begin{table}[hbt]
\centering
\begin{tabular}{ccccc}
\toprule
& $d=15$ & $d=17$ & $d=18$ & $d=19$
\\ \midrule
$b_1$ 
&  	\btenv{$i\bigl(-30+28a-6r_1+4ar_1+(-60+24a-12r_1+8ar_1)t\bigr)^{\frac{1}{2}}$}
& 	\btenv{$ 32 i\bigl( -1938 + 765a + 462r_2 - 181ar_2 - 2(171 - 72a - 67r_2 + 32ar_2)t 
        + 
         2(5844 - 2303a - 1404r_2 + 551ar_2)t^2 + 4(173 - 64a - 229r_2 + 112ar_2)t^3  
         - 
         8(2592 - 1021a - 624r_2 + 245ar_2)t^4 - 16(-39 + 24a - 113r_2 + 56ar_2)t^5 
        + 
         32(348 - 137a - 84r_2 + 33ar_2)t^6 + 512(-2 + a)(1 + r_2)t^7
        \bigr)^{\frac{1}{2}} $}
& 	\btenv{$ i\bigl((a+15)(r_1+5)\bigr)^{\frac{1}{2}}$}
& 	\btenv{$i (2a+1)^{\frac{1}{2}}$} 
\\  \addlinespace[4 pt]
$b_2$ 
& 	\btenv{$2\re\Bigl(\bigl(10+30i)^{\frac{1}{3}}\bigr)\Bigr)$}
& 	\btenv{$2 \re\Bigl( \bigl(17 (44+3 i \sqrt{21})\bigr)^{\frac{1}{3}} \Bigr)$}
& 	\btenv{$ 2\re\Bigl( \bigl(-11+i\sqrt{95}\bigr)^{\frac{1}{3}}\Bigr)$}
& 	\btenv{$2 \re\Bigl( \bigl(38 ( 13 + 3 i \sqrt{5})\bigr)^{\frac{1}{3}}\Bigr)$}
\\  \addlinespace[4 pt]
$b_3$ 
& 	\btenv{$i\sqrt{2a}$}
&
& 	\btenv{$2\re\Bigl(\bigl(\frac{3}{2}(1+i\sqrt{95})\bigr)^{\frac{1}{3}}\Bigr)$}
&
\\  \addlinespace[4 pt]
$b_4$ 
& 	\btenv{$\sqrt{a+2}$}
&
&
&
\\
\bottomrule
\end{tabular}
\caption{\label{tbl:bgens1} $b$ generators for $d=15$, $17$--$19$}
\end{table}

\begin{table}[hbt]
\centering
\begin{tabular}{ccccc}
\toprule
& $d=20$ & $d=21$ & $d=24$ & $d=28$
\\ \midrule
$b_1$ 
&	 \btenv{$i \bigl((a+17)(r_1+17)\bigr)^{\frac{1}{2}}$} 
& 	\btenv{$(9+4 r_1)^{\frac{1}{2}}$} 
&	\btenv{$i\sqrt{a-3}$}
&	\btenv{$i\sqrt{a+1}$}
\\  \addlinespace[4 pt]
$b_2$ 
&	 \btenv{$2 \re\Bigl( \bigl( 10(19+i9\sqrt{119})\bigr)^{\frac{1}{3}}\Bigr)$} 
& 	\btenv{$i \bigl(-33+66a-22r_1+33 a r_1 - (11+8 a+5a r_1)b_1 \bigr)^{\frac{1}{2}}$} 
&	\btenv{$\sqrt{(4+r_1)(3+r_2)}$}
&	\btenv{$i\sqrt{a+5}$}
\\  \addlinespace[4 pt]
$b_3$ 
& 	\btenv{$ \Bigl( \frac{1}{3}\Bigl(\bigl((a+39)r_3r_4+(7a+21)r_3\bigr)t-\bigl((2a+15)r_2-3a\bigr)r_4-3(a+18)r_2+195 \Bigr)^{\frac{1}{2}}$}  
& 	\btenv{$ \bigl(363-66a+66 r_1-33 a r_1 +(55-24 a+7 a r_1) b_1\bigr)^{\frac{1}{2}}$} 
&	\btenv{$4\re\Bigl(\bigl(1+i\sqrt{7}\bigr)^{\frac{1}{3}}\Bigr)$}
&	\btenv{$2\re\Bigl(\bigl(189+21i\sqrt{87}\bigr)^{\frac{1}{3}}\Bigr)$}
\\  \addlinespace[4 pt]
$b_4$ 
&
& 	\btenv{$ \re\Bigl( \bigl( 4 (-4-9 a + i \sqrt{465-72 a})\bigr)^{\frac{1}{3}}\Bigr)$}
&	\btenv{$i\sqrt{(a+1)(r_3+5)}$}
&	\btenv{$i\sqrt{20a+50+(6a+36)r_3}$}
\\
\bottomrule
\end{tabular}
\caption{\label{tbl:bgens2} $b$ generators for $d=20$, $21$, $24$, $28$}
\end{table}

\begin{table}[hbt]
\centering
\begin{tabular}{ccccc}
\toprule
& $d=30$ & $d=35$ & $d=39$ & $d=48$
\\ \midrule
$b_1$ 
&	\btenv{$\frac{1}{2}\bigl( 4(-5r_1r_2 + (4a - 23)r_1)t + 2((-4a + 21)r_2 -8a + 111)\bigr)^{\frac{1}{2}}$}
&	\btenv{$i\sqrt{2a+1}$}
&	$\sqrt{3r_1+18}$
&	$i\sqrt{a-1}$
\\  \addlinespace[4 pt]
$b_2$ 
&	\btenv{$2\re\bigl((1+2 i \sqrt{31})^{\frac{1}{3}}\bigr)$}
&	\btenv{$2\re\bigl( (280+210 i\sqrt{6})^{\frac{1}{3}}\bigr)$}
&	$\sqrt{18r_2+78}$
&	$\sqrt{6+r_1r_2}$
\\  \addlinespace[4 pt]
$b_3$ 
&	\btenv{$2\re\Bigl(\bigl(4+b_2+i\sqrt{48-8b_2-b_2^2}\bigr)^{\frac{1}{3}}\Bigr)$}
&	\btenv{$\sqrt{14(r_1+5)}$}
&	$\sqrt{(4a+15)(r_1+5)}$
&	$\sqrt{6-2r_1r_2+(r_2-r_1)b_2}$
\\  \addlinespace[4 pt]
$b_4$ 
&	\btenv{$2\re\Bigl(\bigl(70+10i\sqrt{31}\bigr)^{\frac{1}{3}}\Bigr)$}
&	\btenv{$\re\Bigl(\bigl(28+84 i\sqrt{3}\bigr)^{\frac{1}{3}}\Bigr)$}
&	\btenv{$\bigl((2a-5)(r_1-2)b_3-(8a-35)(r_1-10)\bigr)^{\frac{1}{2}}$}
&	$\sqrt{2r_3+42}$
\\  \addlinespace[4 pt]
$b_5$ 
&	$\frac{i}{2}\sqrt{2a+18}$
&	$\sqrt{3-r_3}$
&	$\re\bigl( (-676+10140 i\sqrt{3})^{\frac{1}{3}}\bigr)$
&	\btenv{$\bigl((5(a + 1)-(3a - 5)r_3 )b_4 -2(a +5)r_3 +210(3-a)\bigr)^{\frac{1}{2}}$}
\\  \addlinespace[4 pt]
$b_6$ 
&	
&	$\frac{i}{7}\sqrt{245 a-7r_2b_3-49ar_1}$
&	\btenv{$\re\bigl((180-4a+4i\sqrt{6753+90a})^{\frac{1}{3}}\bigr)$}
&	$2\re\bigl( (7+i\sqrt{15})^{\frac{1}{3}}\bigr)$
\\  \addlinespace[4 pt]
$b_7$
&
&
&	$i\sqrt{6a-3}$
&
\\
\bottomrule
\end{tabular}
\caption{\label{tbl:bgens3} $b$ generators for $d=30$, $35$, $39$, $48$}
\end{table}
\end{landscape}
\begin{table}[htb]
\centering
\begin{tabular}{cccc}
\toprule
$d$ & generator & in terms of radicals & minimal polynomial
\\
\midrule
$15$ & $b_2$ & $2\re\Bigl(\bigl(10+30i)\bigr)^{\frac{1}{3}}\Bigr)$ &  $x^3-30x-20$
\\ \addlinespace[4 pt]
$17$ & $b_2$ & $ 2 \re\Bigl( \bigl(17 (44+3 i \sqrt{21})\bigr)^{\frac{1}{3}} \Bigr)$ & $x^3 - 255x - 1496$
\\ \addlinespace[4 pt]
$18$ & $b_2$ & $2\re\Bigl( \bigl(-11+i\sqrt{95}\bigr)^{\frac{1}{3}}\Bigr)$ & $x^3 - 18 x + 22$
\\
 & $b_3$ & $ 2\re\Bigl(\bigl(\frac{3}{2}(1+i\sqrt{95})\bigr)^{\frac{1}{3}}\Bigr)$ & $x^3 - 18 x -3$
 \\ \addlinespace[ 4 pt]
 $19$ & $b_2$ & $ \re\Bigl( \bigl(38 ( 13 + 3 i \sqrt{5})\bigr)^{\frac{1}{3}}\Bigr)$ & $x^3 - 228x- 988$
 \\ \addlinespace[ 4 pt]
 $20$ & $b_2$ & $2 \re\Bigl( \bigl( 10(19+9i\sqrt{119})\bigr)^{\frac{1}{3}}\Bigr)$ & $ x^3 - 300x- 380$
 \\ \addlinespace[ 4 pt]
 $21$ & $b_4$ & $ \re\Bigl( \bigl( 4 (-4-9 a + i \sqrt{465-72 a})\bigr)^{\frac{1}{3}}\Bigr)$  & $x^3 - 21x + 9a + 4$
 \\ \addlinespace[ 4 pt]
 $24$ & $b_3$ & $4\re\Bigl(\bigl(1+i\sqrt{7}\bigr)^{\frac{1}{3}}\Bigr)$ & $x^3-24x-16$
 \\ \addlinespace[ 4 pt]
 $28$ & $b_3$ & $2\re\Bigl(\bigl(189+21i\sqrt{87}\bigr)^{\frac{1}{3}}\Bigr)$ & $x^3-126x-378$
 \\ \addlinespace[ 4 pt]
 $30$ & $b_2$ & $2\re\bigl((1+2 i \sqrt{31})^{\frac{1}{3}}\bigr)$ & $x^3-15x-2$
 \\ \addlinespace[ 4 pt]
 & $b_3$ & $2\re\Bigl(\bigl(4+b_2+i\sqrt{48-8b_2-b_2^2}\bigr)^{\frac{1}{3}}\Bigr)$ & $x^3-12 x-2b_2-8$
 \\ \addlinespace[ 4 pt]
 & $b_4$ & $2\re\Bigl(\bigl(70+10 i \sqrt{31}\bigr)^{\frac{1}{3}}\Bigr)$ & $x^3-60 x-140$
 \\ \addlinespace[ 4 pt]
 $35$ & $b_2$ & $2\re\bigl( (280+210 i\sqrt{6})^{\frac{1}{3}}\bigr)$ & $x^3-210x-560$
 \\ \addlinespace[ 4 pt]
 & $b_4$ & $\re\Bigl(\bigl(28+84 i\sqrt{3}\bigr)^{\frac{1}{3}}\Bigr)$ & $x^3-21x-7$
 \\ \addlinespace[ 4 pt]
 $39$ & $b_5$ & $\re\bigl( (-676+10140 i\sqrt{3})^{\frac{1}{3}}\bigr)$ & $x^3-507x+169$
 \\ \addlinespace[ 4 pt]
 & $b_6$ & $\re\bigl((180-4a+4i\sqrt{6753+90a})^{\frac{1}{3}}\bigr)$ & $x^3-39 x+a-45$
 \\ \addlinespace[ 4 pt]
 $48$ & $b_6$ & $2\re\bigl( (7+i\sqrt{15})^{\frac{1}{3}}\bigr)$ & $x^3-12x-14$
 \\
 \bottomrule
\end{tabular}
\caption{\label{tbl:minpolys}Minimal polynomials for cubic $b$ generators}
\end{table}
\begin{table}[hbt]
\centering
\begin{tabular}{ccccc}
\toprule
multiplet & $\fd{E}_0$ & $\fd{E}_1$  & $\fd{E}$ &  $\deg(\fd{E}/\fd{Q})$ 
\\
\midrule
$15ac$ & $\fd{Q}(a,b_4)$ & $\fd{E}_0(r_1,t,b_1,b_2,b_3)$ & $\fd{E}_1(i)$ & $384$
\\ \addlinespace[1 pt]
$15b$ & $\fd{Q}(a)$ & $\fd{E}_0(r_1,t,b_1,b_2,b_3)$ & $\fd{E}_1(i)$ & $192$
\\ \addlinespace[1 pt]
$15d$ & $\fd{Q}(a)$ & $\fd{E}_0(r_1,t,b_1,b_2)$ & $\fd{E}_1(i)$ & $96$
\\ \addlinespace[1 pt]
\midrule
$17ab$ & $\fd{Q}(a,r_1)$ & $\fd{E}_0(r_2,t,b_1,b_2)$ & $\fd{E}_1(i)$ & $768$
\\ \addlinespace[1 pt]
$17c$ & $\fd{Q}(a)$ & $\fd{E}_0(r_2,t,b_1,b_2)$ & $\fd{E}_1(i)$ & $384$
\\ \addlinespace[1 pt]
\midrule
$18ab$ &  $\fd{Q}(a,r_1)$ & $\fd{E}_0(r_2,t, b_1, b_2, b_3)$ & $\fd{E}_1(i)$ & $864$
\\ \addlinespace[1 pt]
\midrule
$19bc$ & $\fd{Q}(a,r_2)$ & $\fd{E}_0(r_1, t, b_1, b_2)$ & $\fd{E}(i)$ & $864$
\\  \addlinespace[1 pt]
$19a$ & $\fd{Q}(a)$ & $\fd{E}_0(r_1, t, b_1, b_2)$ & $\fd{E}_1(i)$ & $432$
\\  \addlinespace[1 pt]
$19d$ & $\fd{Q}(a)$ & $\fd{E}_0(t,b_1,b_2)$ & $\fd{E}(ir_1)$ & $216$
\\  \addlinespace[1 pt]
$19e$ & $\fd{Q}(a)$ & $\fd{E}_0(t,b_1)$ & $\fd{E}(ir_1)$ & $72$
\\ \addlinespace[1 pt]
\midrule
$20ab$ & $\fd{Q}(a,r_1)$ & $\fd{E}_0 (r_2,r_3,r_4,t,b_1,b_2,b_3)$ & $\fd{E}_1(i)$ & $1536$
\\ \addlinespace[1 pt]
\midrule
$21abcd$ & $\fd{Q}(a,r_1,b_1)$ & $\fd{E}_0(r_2,t,b_2,b_3,b_4)$ & $\fd{E}_1 (i)$ & $1152$
\\  \addlinespace[1 pt]
$21e$ & $\fd{Q}(a)$ & $\fd{E}_0(r_1,r_2,t,b_1,b_2,b_3)$ & $\fd{E}_1(i)$ & $384$
\\  \addlinespace[1 pt]
\midrule
$24ab$ & $\fd{Q}(a,r_3)$  & $\fd{E}_0(r_1,r_2,t,b_1,b_2,b_3,b_4)$ & $\fd{E}_1(i)$ & $1536$
\\  \addlinespace[1 pt]
$24c$ & $\fd{Q}(a)$  & $\fd{E}_0(r_1,r_2,t,b_1,b_2,b_3)$ & $\fd{E}_1(i)$ & $384$
\\  \addlinespace[1 pt]
\midrule
$28ab$ & $\fd{Q}(a,r_3)$ & $\fd{E}_0(r_1,r_2,t,,b_1,b_2,b_3,b_4)$ & $\fd{E}_1(i)$ & $2304$
\\  \addlinespace[1 pt]
$28c$ & $\fd{Q}(a)$ & $\fd{E}_0(r_1,r_2,t,,b_1,b_2,b_3)$ & $\fd{E}_1(i)$ & $576$
\\  \addlinespace[1 pt]
\midrule
$30abc$ & $\fd{Q}(a,b_2)$ & $\fd{E}_0(r_1,r_2,t,b_1,b_3,b_4,b_5)$ & $\fd{E}_1(i)$ & $3456$
\\  \addlinespace[1 pt]
$30d$ & $\fd{Q}(a)$ & $\fd{E}_0(r_1,r_2,t,b_1,b_2,b_4,b_5)$ & $\fd{E}_1(i)$ & $1152$
\\  \addlinespace[1 pt]
\midrule
$35bcdg$ & $\fd{Q}(a,r_3,b_5)$ & $\fd{E}_0(r_1,r_2,b_1,b_2,b_3,b_4,b_6)$ &$\fd{E}_1(i)$ & $4608$
\\  \addlinespace[1 pt]
$35af$ & $\fd{Q}(a,r_3)$ & $\fd{E}_0(r_1,r_2,b_1,b_2,b_3,b_4,b_5b_6)$ &$\fd{E}_1(i)$ & $2304$
\\  \addlinespace[1 pt]
$35e$ & $\fd{Q}(a)$ & $\fd{E}_0(r_1,r_2r_3,b_1,b_2,b_3,b_4,c)$ &$\fd{E}_1(ir_2)$ & $1152$
\\  \addlinespace[1 pt]
$35h$ & $\fd{Q}(a)$ & $\fd{E}_0(r_1,r_2,b_1,b_2,b_3,b_4,b_6)$ &$\fd{E}_1(i)$ & $1152$
\\  \addlinespace[1 pt]
$35i$ & $\fd{Q}(a)$ & $\fd{E}_0(r_1,r_2,b_1,b_2,b_3,b_4)$ &$\fd{E}_1(i)$ & $576$
\\  \addlinespace[1 pt]
$35j$ & $\fd{Q}(a)$ & $\fd{E}_0(r_1,b_1,b_2,b_3,b_4)$ &$\fd{E}_1(ir_2)$ & $288$
\\  \addlinespace[1 pt]
\midrule
$39acde$ & $\fd{Q}(a,r_1,b_1)$ & $\fd{E}_0(r_2,b_2,b_3,b_4,b_5,b_6,b_7)$ & $\fd{E}(ir_3)$ & $4608$
\\  \addlinespace[1 pt]
$39bf$ & $\fd{Q}(a,r_1)$ & $\fd{E}_0(r_2,b_2,b_3,b_4,b_5,b_6,b_7)$ & $\fd{E}(ir_3)$ & $2304$
\\  \addlinespace[1 pt]
$39gh$ & $\fd{Q}(a,r_1)$ & $\fd{E}_0(r_2,b_1,b_2,b_3,b_4,b_5,b_7)$ & $\fd{E}(ir_3)$ & $1536$
\\  \addlinespace[1 pt]
$39ij$ & $\fd{Q}(a,r_1)$ & $\fd{E}_0(r_2,b_2,b_3,b_4,b_5,b_7)$ & $\fd{E}(ir_3)$ & $768$
\\  \addlinespace[1 pt]
\midrule
$48abcd$ & $\fd{Q}(a,r_3,b_4)$ & $\fd{E}_0(r_1,r_2,t,b_1,b_2,b_3,b_5,b_6)$ & $\fd{E}_1(i)$ & $12288$
\\  \addlinespace[1 pt]
$48e$ & $\fd{Q}(a)$ & $\fd{E}_0(r_1,r_2,r_3,t,b_1,b_2,b_3,b_4,b_5)$ & $\fd{E}_1(i)$ & $4096$
\\  \addlinespace[1 pt]
$48f$ & $\fd{Q}(a)$ & $\fd{E}_0(r_1,r_2,t,b_1,b_2,b_3,b_6)$ & $\fd{E}_1(i)$ & $1536$
\\  \addlinespace[1 pt]
$48g$ & $\fd{Q}(a)$ & $\fd{E}_0(r_1,r_2,t,b_1,b_2,b_3)$ & $\fd{E}_1(i)$ & $512$
\\
\bottomrule
\end{tabular}
\caption{\label{tbl:fields} Fields. The exact fiducials for orbits $15d$, $19e$, $24c$, $28c$, $35j$, $48g$ were calculated by Scott and Grassl~\cite{Scott:2010a}.  They are included here for the sake of comparison.The generator $c$ in the field for $35e$ is given by $c=(r_2(ar_1+r_3+2a+1)b_3+42r_1+70)b_5b_6$.}
\end{table}

\begin{landscape}

{\SMALL

\noindent \textbf{Dimension 15}

\vspace{0.1 cm }

\noindent
\parbox[t][4 pt][b]{0.3 cm}{1.}
\fbox{\begin{tabular}[t]{cccccccccccc}
\toprule 
$h$ & $a$ & $r_1$ & $t$  & $b_1$ & $b_2$ & $b_3$ &  $b_4$ & $i$ & order & $g\vpu{-1}_a h g^{-1}_a$
 \\ \addlinespace[2 pt]
\midrule
$g_a$ & $-a$ & $r_1$ & $t$ & $b'_1$ & $b_2$ & $b'_3$ & $b'_4$ & $-i$ & $2$ & $g_a$
\\ \addlinespace[2 pt]
$g_1$ & $a$ & $r_1$ & $t$ & $b_1$ & $b_2$ & $b_3$ & $-b_4$ & $i$ & $2$ & $g_1$
\\ \addlinespace[2 pt]
$g_2$ & $a$ & $-r_1$ & $t'$ & $b''_1$ & $b'_2$ & $b_3$ & $b_4$ & $i$ & $24$ & $g_2^5$
\\ \addlinespace[2 pt]
$g_3$ & $a$ & $r_1$ & $t$ & $b_1$ & $b_2$ & $-b_3$ & $b_4$ & $i$ & $2$ & $g_3$
\\ \addlinespace[2 pt]
$\bar{g}_1$ &  $a$ & $r_1$ & $t$ & $b_1$ & $b_2$ & $b_3$ & $b_4$ & $-i$ & $2$ & $g_c=g_2^{12}g\vpu{12}_3\bar{g}\vpu{12}_1$
\\
\bottomrule
\end{tabular}
\hspace{1 cm}
\parbox[t]{8.5 cm}{ \vspace{0.1 cm} $t'= \cos \frac{7\pi}{15}$ 
\\ 
\vspace{0.05 cm}

$b'_1=\sqrt{30+28a+6r_1+4ar_1+(60+24a+12r_1+8ar_1)t}$
 \\
 \vspace{0.05 cm}
 
$b''_1=i\sqrt{16a+(60-32a-36r_1+16ar_1)t}$
 \\
 \vspace{0.05 cm}
 
 $b'_2=2\re\Bigl(e^{\frac{2\pi i}{3}}(10+30i)^{\frac{1}{3}}\Bigr)$
 \\
 \vspace{0.05 cm}
 
 $b'_3=\sqrt{2a}$
 \\
 \vspace{0.05 cm}
 
 $b'_4=\sqrt{-a+2}$
}
}

\vspace{0.2 cm }

\noindent \parbox[t][4 pt][b]{0.3 cm}{2.}
\fbox{\begin{tabular}[t]{ccccc}
\toprule
orbit & $\gal(\fd{E}/\fd{Q})$ &  $\gal(\fd{E}/\fd{E}_0)$ 
\\
\midrule
$15ac$ & $\la g_a, g_1, g_2, g_3, \bar{g}_1\ra$ &  $\la g_2, g_3, \bar{g}_1\ra$
\\ \addlinespace[2 pt]
$15b$  & $\la g_a,  g_2,g_3, \bar{g}_1\ra$ & $\la g_2, g_3, \bar{g}_1\ra$
\\ \addlinespace[2 pt]
$15d$  & $\la g_a,  g_2, \bar{g}_1\ra$  & $\la g_2, \bar{g}_1\ra$
\\
\bottomrule
\end{tabular}
\hspace{1 cm}
\parbox[t][1.5 cm][c]{9cm}{ $g_1$ interchanges  $15a$, $15c$ and restricts to the identity on $15b$, $15d$;  $g_3$ restricts to the identity on $15d$ }
}
\vspace{0.2 cm }

\noindent \parbox[t][4 pt][b]{0.3 cm}{3.}
\fbox{\begin{tabular}[t]{ccccc}
\toprule
$G_{a1}=G_{c1}$ & $G_{a2}=G_{b2}$ &  $G_{c2}=G_{d2}$ &  $G_{a3}=G_{b3}=G_{c3}$ 
\\ \addlinespace[2 pt]
\midrule
 $\bmt 0 & 1 \\ 1 & 0\emt$ & $\bmt 1 & 12 \\ 3 & 13\emt$ & $\bmt 3 & 14 \\ 1 & 2\emt$ & $\bmt 0 & 4 \\ 11 & 4\emt$
\\ \addlinespace[4 pt]
\bottomrule
\end{tabular}
}
\vspace{0.3 cm}

\noindent \textbf{Dimension 17}

\vspace{0.1 cm }

\noindent
\parbox[t][4 pt][b]{0.3 cm}{1.}
\fbox{\begin{tabular}[t]{cccccccccccc}
\toprule 
$h$ & $a$ & $r_1$ & $r_2$  & $t$ & $b_1$ & $b_2$ &  $i$ & order & $g\vpu{-1}_a h g^{-1}_a$
 \\ \addlinespace[2 pt]
\midrule
$g_a$ & $-a$ & $r_1$ & $r_2$ & $t$ & $b'_1$ & $b_2$   & $i$ & $2$ &  $g_a$
\\ \addlinespace[2 pt]
$g_1$ & $a$ & $-r_1$ & $r_2$ & $t$ & $b_1$ & $b_2$  & $i$  & $2$ & $g_1$
\\ \addlinespace[2 pt]
$g_2$ & $a$ & $r_1$ & $-r_2$ & $t'$ & $b''_1$ & $b'_2$  & $i$ & $96$ & $g_2^{17}$
\\ \addlinespace[2 pt]
$\bar{g}_1$ & $a$ & $r_1$ & $r_2$ & $t$ & $b_1$ & $b_2$ & $-i$ & $2$ & $g_c = g_2^{48} \bar{g}\vpu{48}_1$
\\
\bottomrule
\end{tabular}
\hspace{1 cm}
\parbox[t]{10cm}{ \vspace{0.1 cm} $t'= \sin \frac{20\pi}{17}$ 
\\ 
\vspace{0.05 cm}

$b'_1=32\Bigl(1938 + 765a - 462r_2 - 181ar_2 - 2(-171 - 72a + 67r_2 + 32ar_2)t 
 2(-5844 - 2303a + 1404r_2 + 551ar_2)t^2 + 4(-173 - 64a + 229r_2 + 112ar_2)
  t^3 - 8(-2592 - 1021a + 624r_2 + 245ar_2)t^4 - 
 16(39 + 24a + 113r_2 + 56ar_2)t^5+ 32(-348 - 137a + 84r_2 + 33ar_2)t^6 + 
 512(2 + a)(1 + r_2)t^7\Bigr)^{\frac{1}{2}}$
 \\
 \vspace{0.05 cm}
 
$b''_1=32 i \Bigl(\frac{1}{2}(-204 + 119a + 60r_2 - 33ar_2) - 8(-39 + 20a - 3r_2)t + 
 16(-81 + 25a - 39r_2 + 17ar_2)t^2 + 4(-491 + 240a - 51r_2)t^3  - 
 8(-600 + 211a - 216r_2 + 91ar_2)t^4 - 16(-243 + 112a - 27r_2)t^5+ 
 96(-36 + 13a - 12r_2 + 5ar_2)t^6 + 256(-9 + 4a - r_2)t^7\Bigr)^{\frac{1}{2}}$
 \\
 \vspace{0.05 cm}
 
 $b'_2=2 \re\Bigl(e^{\frac{4 \pi i}{3}} \bigl(17(44+3 i \sqrt{21}) \bigr)^{\frac{1}{3}}\Bigr)$
}
}

\vspace{0.2 cm }

\noindent \parbox[t][4 pt][b]{0.3 cm}{2.}
\fbox{\begin{tabular}[t]{ccccc}
\toprule
orbit & $\gal(\fd{E}/\fd{Q})$ &  $\gal(\fd{E}/\fd{E}_0)$ 
\\
\midrule
$17ab$ & $\la g_a, g_1, g_2, \bar{g}_1\ra$ & $\la g_2, \bar{g}_1\ra$ 
\\ \addlinespace[2 pt]
$17c$  & $\la g_a,  g_2, \bar{g}_1\ra$ & $\la g_2, \bar{g}_1\ra$
\\
\bottomrule
\end{tabular}
\hspace{1 cm}
\parbox[t][1.5 cm][c]{9cm}{ $g_1$ interchanges orbits $17a$, $17b$ and restricts to the identity on $17c$.}
}
\vspace{0.2 cm }

\noindent \parbox[t][4 pt][b]{0.3 cm}{3.}
\fbox{\begin{tabular}[t]{ccccc}
\toprule
$G_{a1}$ & $G_{b1}$ &  $G_{a2}$ &  $G_{b2}$ &  $G_{c2}$ 
\\ \addlinespace[2 pt]
\midrule
 $\bmt 2 & 7 \\ 9 & 15\emt$ & $\bmt 15 & 10 \\ 8 & 2 \emt$ &  $\bmt 1 & 11 \\ 6 & 12\emt$ &
 $\bmt 6 & 16 \\ 1 & 5 \emt$ &
  $\bmt 1 & 11 \\ 6 & 12 \emt$ 
\\ \addlinespace[4 pt]
\bottomrule
\end{tabular}
}
\vspace{0.3 cm}

\noindent \textbf{Dimension 18}

\vspace{0.1 cm }

\noindent
\parbox[t][4 pt][b]{0.3 cm}{1.}
\fbox{\begin{tabular}[t]{cccccccccccc}
\toprule 
$h$ & $a$ & $r_1$ & $r_2$  & $t$ & $b_1$ & $b_2$ & $b_3$ & $i$ & order & $g\vpu{-1}_a h g^{-1}_a$
 \\ \addlinespace[2 pt]
\midrule
$g_a$ & $-a$ & $r_1$ & $r_2$ & $t$ & $b'_1$ & $b_2$  & $b_3$ & $i$ & $2$ &  $g_a$
\\ \addlinespace[2 pt]
$g_1$ & $a$ & $-r_1$ & $r_2$ & $t$ & $b''_1$ & $b_2$ & $b'_3$ & $i$  & $12$ & $g^{11}_1$
\\ \addlinespace[2 pt]
$g_2$ & $a$ & $r_1$ & $-r_2$ & $-t$ & $b_1$ & $b'_2$ & $b_3$ & $i$ & $6$ & $g_1^6 g_2^5$
\\ \addlinespace[2 pt]
$g_3$ & $a$ & $r_1$ & $r_2$ & $t'$ & $b_1$ & $b_2$ & $b_3$ & $i$ & $3$ & $g_3$
\\ \addlinespace[2 pt]
$\bar{g}_1$ & $a$ & $r_1$ & $r_2$ & $t$ & $b_1$ & $b_2$ & $b_3$ & $-i$ & $2$ & $g_c =g_1^3 \bar{g}\vpu{3}_1$
\\ \addlinespace[2 pt]
\bottomrule
\end{tabular}
\hspace{1 cm}
\parbox[t]{9.7 cm}{ \vspace{0.1 cm} 
$t'=\cos \frac{11 \pi}{18}$
\\
\vspace{0.05 cm}

$b'_1=-\bigl((a-15)(5+r_1)\bigr)^{\frac{1}{2}}$
\\
\vspace{0.05 cm}

$b''_1=i \bigl(a+15)(5-r_1)\bigr)^{\frac{1}{2}}$
\\
\vspace{0.05 cm}

$b'_2=2 \re\Bigl(e^{\frac{4 \pi i}{3}} \bigl( -11 + i \sqrt{95}\bigl)^{\frac{1}{3}}\Bigr)$
\\
\vspace{0.05 cm}

$b'_3=2 \re\Bigl( e^{\frac{4 \pi i}{3}} \bigl(\frac{3}{2} (1+i\sqrt{95})^{\frac{1}{3}}\bigr)\bigr)$
}
}

\vspace{0.2 cm }

\noindent \parbox[t][4 pt][b]{0.3 cm}{2.}
\fbox{\begin{tabular}[t]{ccccc}
\toprule
orbit & $\gal(\fd{E}/\fd{Q})$ & $\gal(\fd{E}/\fd{E}_0)$
\\
\midrule
$18ab$ & $\la g\vpu{2}_a, g\vpu{2}_1, g\vpu{2}_2, g\vpu{2}_3, \bar{g}\vpu{2}_1\ra$  & $\la g_1^2, g\vpu{2}_2, g\vpu{2}_3 , \bar{g}\vpu{2}_1\ra$
\\ 
\bottomrule
\end{tabular}
\hspace{1 cm}
\parbox[t][0.9 cm][c]{9.3 cm}{ $g_1$ interchanges orbits $18a$, $18b$.}
}

\vspace{0.5 cm }

\noindent \parbox[t][4 pt][b]{0.3 cm}{3.}
\fbox{\begin{tabular}[t]{ccccc}
\toprule
 $G_{a1}$ & $G_{b1}$ &  $G_{a2}$ &  $G_{b2}$ & $G_{a3}$, $G_{b3}$ 
\\ \addlinespace[2 pt]
\midrule
$\bmt 1 & 8 \\ 9 & 35 \emt$ & $\bmt 4 & 3 \\ 7 & 32\emt$ & $\bmt 9 & 17 \\ 19 & 26 \emt$ &   $\bmt 1 & 9 \\ 27 &10 \emt$  & $\bmt 0 & 5 \\ 31 & 5  \emt$ 
\\ \addlinespace[4 pt]
\bottomrule
\end{tabular}
}

\vspace{0.5 cm}

\noindent \textbf{Dimension 19}

\vspace{0.1 cm }

\noindent
\parbox[t][4 pt][b]{0.3 cm}{1.}
\fbox{\begin{tabular}[t]{ccccccccccc}
\toprule 
$h$ & $a$ & $r_1$ & $r_2$  & $t$ & $b_1$ & $b_2$ & $ir_1$ & order & $g\vpu{-1}_a h g^{-1}_a$ 
\\  \addlinespace[2 pt]
\midrule
$g_a$ & $-a$ & $r_1$ & $r_2$ & $t$ & $b'_1$ & $b_2$  & $ir_1$ & $2$ &  $g_a$
\\ \addlinespace[2 pt]
$g_1$ & $a$ & $r_1$ & $-r_2$ & $t$ & $b_1$ & $b_2$ & $ir_1$  & $2$ & $g_1$
\\ \addlinespace[2 pt]
$g_2$ & $a$ & $r_1$ & $r_2$ & $t'$ & $-b_1$ & $b_2$ & $ir_1$ & $18$ & $g_2$
\\ \addlinespace[2 pt]
$g_3$ & $a$ & $r_1$ & $r_2$ & $t$ & $b_1$ & $b'_2$ & $ir_1$ & $3$ & $g^2_3$
\\ \addlinespace[2 pt]
$g_4$ & $a$ & $-r_1$ & $r_2$ & $t$ & $b_1$ & $b_2$ & $ir_1$ & $2$ & $g_4$
\\ \addlinespace[2 pt]
$\bar{g}_1$ & $a$ & $r_1$ & $r_2$ & $t$ & $b_1$ & $b_2$ & $-ir_1$ & $2$ & $g_c = g_2^9 \bar{g}\vpu{9}_1$
\\
\bottomrule
\end{tabular}
\hspace{1 cm}
\parbox[t]{10.25 cm}{ \vspace{0.1 cm} 
$t'=\cos \frac{3 \pi}{19}$
\\
\vspace{0.05 cm}

$b'_1=(2a-1)^{\frac{1}{2}}$
\\
\vspace{0.05 cm}

$b'_2=2 \re\Bigl(e^{\frac{4 \pi i}{3}} \bigl(38 ( 13 + 3 i \sqrt{5})\bigr)^{\frac{1}{3}}\Bigr)$
}
}

\vspace{0.2 cm }

\noindent \parbox[t][4 pt][b]{0.3 cm}{2.}
\fbox{\begin{tabular}[t]{ccccc}
\toprule
orbit & $\gal(\fd{E}/\fd{Q})$ & $\gal(\fd{E}/\fd{E}_0)$ 
\\
\midrule
$19bc$ & $\la g_a, g_1,g_2,g_3,g_4,\bar{g}_1 \ra$ & $\la g_2, g_3, g_4, \bar{g}_1\ra$ 
\\ \addlinespace[2 pt]
$19a$ & $\la g_a, g_2,g_3,g_4,\bar{g}_1 \ra$ & $\la g_2, g_3, g_4, \bar{g}_1\ra$ 
\\ \addlinespace[2 pt]
$19d$ & $\la g_a, g_2, g_3, \bar{g}_1\ra$ & $\la g_2 , g_3, \bar{g}_1\ra$ 
\\ \addlinespace[2 pt]
$19e$ & $\la g_a, g_2, \bar{g}_1\ra$  & $\la g_2, \bar{g}_1\ra$
\\
\bottomrule
\end{tabular}
\hspace{1 cm}
\parbox[t][2 cm][c]{8.75 cm}{ $g_1$ interchanges orbits $19b$, $19c$ and restricts to the identity on $19a$, $19d$, $19e$.  $g_3$ restricts  to the identity on $19e$;  $g_4$ restricts to the identity on $19d$, $19e$. }
}

\vspace{0.5 cm }

\noindent \parbox[t][4 pt][b]{0.3 cm}{3.}
\fbox{\begin{tabular}[t]{cccccccc}
\toprule
$G_{b1}$ & $G_{c1}$ & $G_{a2}$, $G_{b2}$, $G_{c2}$, $G_{d2}$, $G_{e2}$ & $G_{a3}$ &  $G_{b3}$, $G_{c3}$ & $G_{d3}$ & $G_{a4}$ &   $G_{b4}$, $G_{c4}$
\\ \addlinespace[2 pt]
\midrule
 $\bmt 2 & 5 \\ 14 & 7\emt$ & $\bmt 7 & 14 \\ 5 & 2\emt$  &
 $\bmt 15 & 0 \\ 0 & 15\emt$ & $\bmt 7 & 14 \\ 5 & 2\emt$ &  $\bmt 2 & 5 \\ 14 & 7\emt$ & $\bmt 7 & 14 \\ 5 & 2\emt$ & $\bmt 10 & 14 \\ 5 & 5 \emt$ &
 $\bmt 5 & 5 \\ 14 & 10\emt$ 
\\ \addlinespace[4 pt]
\bottomrule
\end{tabular}
}

\vspace{0.5 cm}

\noindent \textbf{Dimension 20}

\vspace{0.1 cm }

\noindent
\parbox[t][4 pt][b]{0.3 cm}{1.}
\fbox{\begin{tabular}[t]{ccccccccccccc}
\toprule 
$h$ & $a$ & $r_1$ & $r_2$ & $r_3$ & $r_4$ & $t$ & $b_1$ & $b_2$ & $b_3$ & $i$ & order & $g\vpu{-1}_a h g^{-1}_a$ \\
 \midrule
 $g_a$ & $-a$ & $r_1$ & $r_2$ & $r_3$ & $r_4$ & $t$ & $b'_1$ & $b_2$ & $b'_3$ & $i$ & 2 & $g_a$
 \\ \addlinespace[2 pt]
 $g_1$ & $a$ & $-r_1$ & $r_2$ & $r_3$ & $r_4$ & $t$ & $b''_1$ & $b_2$ & $b_3$ & $i$  & 4 & $g_1^3$
 \\ \addlinespace[2 pt]
 $g_2$ & $a$ & $r_1$ & $r_2$ & $r_3$ & $-r_4$ & $t'$ & $b_1$ & $b'_2$ & $b''_3$ & $i$ &24 & $g_2^5$ 
 \\ \addlinespace[2 pt]
 $g_3$ & $a$ & $r_1$ & $-r_2$ & $r_3$ & $r_4$ & $t''$ & $b_1$ & $b_2$ & $b_3$ & $i$ & 2 & $g_2^{12} g\vpu{12}_3$
 \\ \addlinespace[2 pt]
 $g_4$ & $a$ & $r_1$ & $r_2$ & $-r_3$ & $r_4$ & $-t$ & $b_1$ & $b_2$ & $b_3$ & $i$ & 2 & $g_4$
 \\ \addlinespace[2 pt]
 $\bar{g}_1$ & $a$ & $r_1$ & $r_2$ & $r_3$ & $r_4$ & $t$ & $b_1$ & $b_2$ & $b_3$ & $-i$ & 2 & $g\vpu{2}_c = g_1^2\bar{g}\vpu{2}_1$ 
 \\ \addlinespace[2 pt]
 \bottomrule
 \end{tabular}
\hspace{1 cm}
\parbox[t]{8 cm}{ \vspace{0.1 cm} 
$t'=\cos \frac{17 \pi}{20}$
\\  
\vspace{0.05 cm}

$t''=\cos \frac{11\pi}{20}$
\\ 
\vspace{0.05 cm}

$b'_1=\bigl((a-17)(17+r_1)\bigr)^{\frac{1}{2}}$
\\ 
\vspace{0.05 cm}

$b''_1=i\bigl((a+17)(17-r_1)\bigr)^{\frac{1}{2}}$
\\  
\vspace{0.05 cm}

$b'_2=2\re\bigl(10e^{\frac{2\pi i}{3}}(19+i\sqrt{119})^{\frac{1}{3}}\bigr)$ 
\\  
\vspace{0.05 cm}

$b'_3=\Bigl(\frac{1}{3}\Bigl(\bigl((-a+39)r_3r_4+(-7a+21)r_3\bigr)t-\bigl((-2a+15)r_2+3a\bigr)r_4$ \\ $\phantom{b'_3= \Bigl(} -3(-a+18)r_2+195\Bigr)\Bigr)^{\frac{1}{2}}$
\\  
\vspace{0.05 cm}

$b''_3=-\Bigl(\frac{1}{3}\Bigl(\bigl(-(a+39)r_3r_4+(7a+21)r_3\bigr)t'+\bigl((2a+15)r_2-3a\bigr)r_4$ \\ $\phantom{b''_3=\Bigl(} -3(a+18)r_2+195\Bigr)\Bigr)^{\frac{1}{2}}$
}
}

\vspace{0.2 cm }

\noindent \parbox[t][4 pt][b]{0.3 cm}{2.}
\fbox{\begin{tabular}[t]{ccccc}
\toprule
orbit & $\gal(\fd{E}/\fd{Q})$ &  $\gal(\fd{E}/\fd{E}_0)$ 
\\
\midrule \addlinespace[2 pt]
$20ab$ & $\la g_a, g_1, g_2, g_3, g_4, \bar{g}_1\ra$ & $\la g_1^2, g\vpu{2}_2, g\vpu{2}_3, g\vpu{2}_4,\bar{g}\vpu{2}_1\ra$
\\ \addlinespace[2 pt]
\bottomrule
\end{tabular}
\hspace{1 cm}
\parbox[t][1  cm][c]{8.1 cm}{ $g_1$ interchanges orbits $20a$, $20b$.}
}

\vspace{0.5 cm }

\noindent \parbox[t][4 pt][b]{0.3 cm}{3.}
\fbox{\begin{tabular}[t]{ccccc}
\toprule
$G_{a1}$ & $G_{b1}$ & $G_{a2}$, $G_{b2}$ & $G_{a3}$, $G_{b3}$ &  $G_{a4}$, $G_{b4}$
\\ \addlinespace[2 pt]
\midrule
$\bmt 1 & 24 \\ 16 & 25 \emt$ & $\bmt 15 & 24 \\ 16 & 39\emt$ & $\bmt 1 & 32 \\ 8 & 33 \emt$ & $\bmt 10 & 31 \\ 9 & 1 \emt$ & $\bmt 1 & 20 \\ 20 & 21 \emt$
\\ \addlinespace[4 pt]
\bottomrule
\end{tabular}
}

\vspace{0.5  cm}

\noindent \textbf{Dimension 21}

\vspace{0.1 cm }

\noindent
\parbox[t][4 pt][b]{0.3 cm}{1.}
\fbox{\begin{tabular}[t]{ccccccccccccc}
\toprule 
$h$ & $a$ & $r_1$ & $r_2$  & $t$ & $b_1$ & $b_2$ & $b_3$ & $b_4$ & $i$ & order & $g\vpu{-1}_a h g^{-1}_a$ \\
\midrule
$g_a$ & $-a$ & $r_1$ & $r_2$ & $t$ & $b_1$ & $b'_2$  & $b'_3$ & $b'_4$ & $i$ & $2$ &  $g_a$
\\
$g_1$ & $a$ & $-r_1$ & $r_2$ & $t'$ & $b'_1$ & $b''_2$ & $b''_3$ & $b_4$ & $i$ & $24$ & $g_1^{19} g_4^2$
\\
$g_2$ & $a$ & $r_1$ & $-r_2$ & $t''$ & $b_1$ & $b_2$ & $b_3$ & $b_4$ & $i$ & $2$ & $g_2 g_3$
\\
$g_3$ & $a$ & $r_1$ & $r_2$ & $t$ & $b_1$ & $b_2$ & $-b_3$ & $b_4$ & $i$ & $2$ & $g_3$
\\
$g_4$ & $a$ & $r_1$ & $r_2$ & $t$ & $b_1$ & $b_2$ & $b_3$ & $b''_4$ & $i$ & $2$ & $g_4^2$
\\
$\bar{g}_1$ & $a$ & $r_1$ & $r_2$ & $t$ & $b_1$ & $b_2$ & $b_3$ & $b_4$ & $-i$ & $2$ & $g\vpu{12}_c = g_1^{12} g\vpu{12}_3 \bar{g}\vpu{12}_1$
\\
\bottomrule
\end{tabular}
\hspace{1 cm}
\parbox[t]{8.35 cm}{ \vspace{0.1 cm} 
$t' = \cos \frac{19 \pi}{21}$
\\ 
\vspace{0.05 cm}

$t'' = \cos \frac{13 \pi}{21}$
\\ 
\vspace{0.05 cm}

$b'_1 = -(9-4r_1)^{\frac{1}{2}}$
\\ 
\vspace{0.05 cm}

$b'_2 = -\bigl(33+66a+22r_1+33a r_1+(11-8a-5ar_1)b_1\bigr)^{\frac{1}{2}}$
\\ 
\vspace{0.05 cm}

$b''_2 = i\bigl(-33+66a+22r_1-33ar_1 - (77+4a-44r_1-3ar_1)b_1\bigr)^{\frac{1}{2}}$
\\ 
\vspace{0.05 cm}

$b'_3 = \bigl(363+66a+66r_1+33ar_1 + (55+24a-7a r_1)b_1\bigr)^{\frac{1}{2}}$
\\ 
\vspace{0.05 cm}

$b''_3 = -\bigl(363-66a-66r_1+33a r_1-(33-20a-44r_1+15a r_1)b_1\bigr)^{\frac{1}{2}}$
\\  
\vspace{0.05 cm}

$b'_4 = \re\Bigl( e^{\frac{4 \pi i}{3}} \bigl( 4 (-4+9a+i\sqrt{465+72 a})\bigr)^{\frac{1}{3}}\Bigr)$
\\   
\vspace{0.05 cm}

$b''_4= \re\Bigl( e^{\frac{2 \pi i}{3}} \bigl( 4 (-4-9a+i\sqrt{465-72 a})\bigr)^{\frac{1}{3}}\Bigr)$
}
}

\vspace{0.2 cm }

\noindent \parbox[t][4 pt][b]{0.3 cm}{2.}
\fbox{\begin{tabular}[t]{ccccc}
\toprule
orbit & $\gal(\fd{E}/\fd{Q})$ & $\gal(\fd{E}/\fd{E}_0)$  
\\
\midrule \addlinespace[2 pt]
$21abcd$ & $\la g_a, g_1, g_2, g_3, g_4, \bar{g}_1\ra$ & $\la g_1^4, g\vpu{2}_2, g\vpu{2}_3, g\vpu{2}_4,\bar{g}\vpu{2}_1\ra$
\\ \addlinespace[2 pt]
$21e$ & $\la g_a, g_1, g_2, g_3, \bar{g}_1\ra$ &  $\la g_1, g_2, g_3, \bar{g}_1\ra$
\\ \addlinespace[2 pt]
\bottomrule
\end{tabular}
\hspace{1 cm}
\parbox[t][1  cm][c]{7.7 cm}{ $g_1$ cycles  orbits $21abcd$ in the order $21a\to 21d \to 21b \to 21c \to 21a \to \dots$;  $g_4$ restricts to the identity on $21e$.}
}


\vspace{0.5 cm }

\noindent \parbox[t][4 pt][b]{0.3 cm}{3.}
\fbox{\begin{tabular}[t]{cccccccccc}
\toprule
$G_{a1}$ & $G_{b1}$ & $G_{c1}$ & $G_{d1}$ & $G_{e1}$ & $G_{a2}$, $G_{b2}$, $G_{c2}$, $G_{d2}$ & $G_{e2}$ &  $G_{a3}, G_{b3}, G_{c3}, G_{d3}$ & $G_{e3}$ & $G_{a4}$, $G_{b4}$, $G_{c4}$, $G_{d4}$
\\ \addlinespace[2 pt]
\midrule
$\bmt 2 & 9 \\ 12 & 11 \emt$ & $\bmt 11 & 12 \\ 9 & 2 \emt$ & $\bmt 2 & 3 \\ 18 & 5 \emt$ & $\bmt 11 & 19 \\ 2 & 9 \emt$ & $\bmt 4 & 7 \\ 7 & 18 \emt$ &
$\bmt 10 & 15 \\ 6 & 4 \emt$ & $\bmt 7 & 9 \\ 18 & 19\emt$ & $\bmt 0 & 8 \\ 13 & 8 \emt$ & $\bmt 13 & 0 \\ 0 & 13 \emt$ & $\bmt 6 & 8 \\ 13 & 14 \emt$
\\ \addlinespace[4 pt]
\bottomrule
\end{tabular}
}

\vspace{0.5 cm}

\noindent \textbf{Dimension 24}

\vspace{0.1 cm }

\noindent
\parbox[t][4 pt][b]{0.3 cm}{1.}
\fbox{\begin{tabular}[t]{ccccccccccccc}
\toprule 
$h$ & $a$ & $r_1$ & $r_2$  & $r_3$ & $t$ & $b_1$ & $b_2$ & $b_3$ & $b_4$ & $i$ & order & $g\vpu{-1}_a h g^{-1}_a$ \\
\midrule
$g_a$ & $-a$ & $r_1$ & $r_2$ & $r_3$ & $t$ & $b'_1$ & $b_2$ & $b_3$ & $b'_4$ & $i$ & $2$ & $g_a$
\\
$g_1$  & $a$ & $r_1$ & $r_2$ & $-r_3$ & $t$ & $b_1$ & $b_2$ & $b_3$ & $b''_4$ & $i$ & $4$ & $g^3_1$
\\
$g_2$ & $a$ & $r_1$ & $-r_2$ & $r_3$ & $t'$ & $b_1$ & $b'_2$ & $b'_3$ & $b_4$ & $i$ & $12$ & $g^5_2g\vpu{5}_4$
\\
$g_3$ &  $a$ & $-r_1$ & $r_2$ & $r_3$ & $t''$ & $b_1$ & $b''_2$ & $b_3$ & $b_4$ & $i$ & $4$ & $g^6_2g\vpu{6}_3$
\\
$g_4$ & $a$ & $r_1$ & $r_2$ & $r_3$ & $t$ & $-b_1$ & $b_2$ & $b_3$ & $b_4$ & $i$ & $2$ & $g_4$
\\
$\bar{g}_1$ & $a$ & $r_1$ & $r_2$ & $r_3$ & $t$ & $b_1$ & $b_2$ & $b_3$ & $b_4$ & $-i$ & $2$ & $g_c =g^2_1g\vpu{2}_4\bar{g}\vpu{2}_1$
\\
\bottomrule
\end{tabular}
\hspace{1 cm}
\parbox[t]{4.5 cm}{ \vspace{0.1 cm} 
$t' = \cos \frac{7\pi}{24}$
\\ 
\vspace{0.05 cm}

$t'' = \cos \frac{11\pi}{24} $
\\ 
\vspace{0.05 cm}

$b'_1 =\sqrt{a+3}$
\\ 
\vspace{0.05 cm}

$b'_2 = \sqrt{(4+r_1)(3-r_2)}$
\\ 
\vspace{0.05 cm}

$b''_2 =\sqrt{(4-r_1)(3+r_2)}$
\\ 
\vspace{0.05 cm}

$b'_3 = 4 \re\Bigl( e^{\frac{2\pi i}{3}} (1+i\sqrt{7})^{\frac{1}{3}}\Bigr) $
\\  
\vspace{0.05 cm}

$b'_4 =- \sqrt{(a-1)(r_3+5)}$
\\   
\vspace{0.05 cm}

$b''_4= i\sqrt{(a+1)(-r_3+5)}$
}
}

\vspace{0.2 cm }

\noindent \parbox[t][4 pt][b]{0.3 cm}{2.}
\fbox{\begin{tabular}[t]{ccc}
\toprule
orbit & $\gal(\fd{E}/\fd{Q})$ &  $\gal(\fd{E}/\fd{E}_0)$ 
\\
\midrule \addlinespace[2 pt]
$24ab$ & $\la g_a,g_1,g_2,g_3,g_4,\bar{g}_1\ra$ & $\la g_1^2,g\vpu{2}_2,g\vpu{2}_3,g\vpu{2}_4,\bar{g}\vpu{2}_1\ra$ 
\\ \addlinespace[2 pt]
$24c$ & $\la g_a,g_2,g_3,g_4,\bar{g}_1\ra$ & $\la g_2,g_3,g_4,\bar{g}_1\ra$
\\ \addlinespace[2 pt]
\bottomrule
\end{tabular}
\hspace{1 cm}
\parbox[t][1  cm][c]{8.9 cm}{ $g_1$ interchanges $24a$, $24b$ and restricts to the identity on $24c$.}
}


\vspace{0.5 cm }

\noindent \parbox[t][4 pt][b]{0.3 cm}{3.}
\fbox{\begin{tabular}[t]{cccccccc}
\toprule
$G_{a1}$ & $G_{b1}$ & $G_{a2}$ & $G_{b2}=G_{c2}$ & $G_{a3}$ & $G_{b3}=G_{c3}$ & $G_{a4}=G_{b4}$ & $G_{c4}$
\\ \addlinespace[2 pt]
\midrule
$\bmt 9 & 23 \\ 32 & 39\emt$ & $\bmt 3 & 5 \\ 8 & 45 \emt$ & $\bmt 11 & 46 \\ 2 & 9 \emt$ & $\bmt 2 & 37 \\ 11 & 39 \emt$
& $\bmt 17 & 36 \\ 12 & 5 \emt$ & $\bmt 5 & 12 \\ 36 & 17\emt$ & $\bmt 5 & 24 \\ 24 & 29\emt$ & $\bmt 0 & 25 \\ 23 & 25\emt$
\\ \addlinespace[4 pt]
\bottomrule
\end{tabular}
}

\vspace{0.5  cm}

\noindent \textbf{Dimension 28}

\vspace{0.1 cm }

\noindent
\parbox[t][4 pt][b]{0.3 cm}{1.}
\fbox{\begin{tabular}[t]{ccccccccccccc}
\toprule 
$h$ & $a$ & $r_1$ & $r_2$  & $r_3$ & $t$ & $b_1$ & $b_2$ & $b_3$ & $b_4$ & $i$ & order & $g\vpu{-1}_a h g^{-1}_a$
 \\
\midrule
$g_a$ & $-a$ & $r_1$ & $r_2$ & $r_3$ & $t$ & $b'_1$ & $b'_2$ & $b_3$ & $b'_4$ & $i$ & $2$ & $g_a$
\\
$g_1$  & $a$ & $r_1$ & $r_2$ & $-r_3$ & $t$ & $b_1$ & $b_2$ & $b_3$ & $b''_4$ & $i$ & $4$ & $g^3_1$
\\
$g_2$ & $a$ & $-r_1$ & $r_2$ & $r_3$ & $t'$ & $b_1$ & $b_2$ & $b_3$ & $b_4$ & $i$ & $6$ & $g_2$
\\
$g_3$ &  $a$ & $r_1$ & $r_2$ & $r_3$ & $t$ & $b_1$ & $-b_2$ & $b'_3$ & $b_4$ & $i$ & $6$ & $g^5_3$
\\
$g_4$ & $a$ & $r_1$ & $-r_2$ & $r_3$ & $t''$ & $b_1$ & $b_2$ & $b_3$ & $b_4$ & $i$ & $2$ & $g^2_1g^3_3g\vpu{3}_4g\vpu{3}_5$
\\
$g_5$ & $a$ & $r_1$ & $r_2$ & $r_3$ & $t$ & $-b_1$ & $-b_2$ & $b_3$ & $b_4$ & $i$ & $2$ & $g_5$
\\
$\bar{g}_1$ & $a$ & $r_1$ & $r_2$ & $r_3$ & $t$ & $b_1$ & $b_2$ & $b_3$ & $b_4$ & $-i$ & $2$ & $g\vpu{2}_c = g^2_1g\vpu{2}_5\bar{g}\vpu{2}_1$
\\
\bottomrule
\end{tabular}
\hspace{1 cm}
\parbox[t]{4.5 cm}{ \vspace{0.1 cm} 
$t' =\cos \frac{3\pi}{28} $
\\ 
\vspace{0.05 cm}

$t'' = \cos \frac{15\pi}{28}$
\\ 
\vspace{0.05 cm}

$b'_1 =\sqrt{a-1}$
\\ 
\vspace{0.05 cm}

$b'_2 =\sqrt{a-5}$
\\ 
\vspace{0.05 cm}

$b'_3 =  2\re\Bigl(e^{\frac{2\pi i}{3}} (189+21i \sqrt{87})^{\frac{1}{3}}\Bigr)$
\\  
\vspace{0.05 cm}

$b'_4 =-\sqrt{20a-50+(6a-36)r_3}$
\\   
\vspace{0.05 cm}

$b''_4=i\sqrt{20a+50-(6a+36)r_3}$
}
}

\vspace{0.2 cm }

\noindent \parbox[t][4 pt][b]{0.3 cm}{2.}
\fbox{\begin{tabular}[t]{ccc}
\toprule
orbit & $\gal(\fd{E}/\fd{Q})$ &  $\gal(\fd{E}/\fd{E}_0)$ 
\\
\midrule \addlinespace[2 pt]
$28ab$ & $\la g_a,g_1,g_2,g_3,g_4,g_5,bar{g}_1\ra$ & $\la g_1^2,g\vpu{2}_2,g\vpu{2}_3,g\vpu{2}_4,g\vpu{2}_5,\bar{g}\vpu{2}_1\ra$ 
\\ \addlinespace[2 pt]
$28c$ & $\la g_a,g_2,g_3,g_4,g_5,\bar{g}_1\ra$ & $\la g_2,g_3,g_4,g_5,\bar{g}_1\ra$
\\ \addlinespace[2 pt]
\bottomrule
\end{tabular}
\hspace{1 cm}
\parbox[t][1  cm][c]{7.9 cm}{ $g_1$ interchanges $28a$, $28b$ and restricts to the identity on $28c$.}
}


\vspace{0.5 cm }

\noindent \parbox[t][4 pt][b]{0.3 cm}{3.}
\fbox{\begin{tabular}[t]{cccccccccc}
\toprule
$G_{a1}$ & $G_{b1}$ & $G_{a2}=G_{b2}$ & $G_{c2}$ & $G_{a3}=G_{b3}$ & $G_{c3}$& $G_{a4}=G_{b4}$ & $G_{c4}$& $G_{a5}=G_{b5}$ & $G_{c5}$
\\ \addlinespace[2 pt]
\midrule
$\bmt 17 & 50 \\ 6 & 11\emt$ & $\bmt 0 & 29 \\ 27 & 29 \emt$ & $\bmt 13 & 30 \\ 26 & 43\emt$ & $\bmt 9 & 19 \\ 37 & 28\emt$&
$\bmt 8 & 49 \\ 7 &1 \emt$ & $\bmt 1 & 7\\ 49 & 8\emt$ & $\bmt 14 & 55 \\ 1 & 13 \emt$ & $\bmt 1 & 42 \\ 14 & 43\emt$ &
$\bmt 17 & 50 \\ 6 & 11\emt$ & $\bmt 0 & 29 \\ 27 & 29\emt$
\\ \addlinespace[4 pt]
\bottomrule
\end{tabular}
}

\vspace{0.5 cm}

\noindent \textbf{Dimension 30}

\vspace{0.1 cm }

\noindent
\parbox[t][4 pt][b]{0.3 cm}{1.}
\fbox{\begin{tabular}[t]{ccccccccccccc}
\toprule 
$h$ & $a$ & $r_1$ & $r_2$  & $t$ & $b_1$ & $b_2$ & $b_3$ & $b_4$ &  $b_5$ & $i$ & order & $g\vpu{-1}_a h g^{-1}_a$
 \\
\midrule
$g_a$ & $-a$ & $r_1$ & $r_2$ & $t$ & $b'_1$ & $b_2$ & $b_3$ & $b_4$ & $b'_5$ & $i$ & $2$ & $g_a$
\\
$g_1$  & $a$ & $r_1$ & $r_2$ & $t$ & $b_1$ & $b'_2$ & $b'_3$ & $b_4$ & $b_5$ & $i$ & $9$ & $g^8_1$
\\
$g_2$ & $a$ & $r_1$ & $-r_2$ & $t'$ & $b''_1$ & $b_2$ & $b_3$ & $b_4$ & $b_5$ & $i$ & $8$ & $g^5_2$
\\
$g_3$ &  $a$ & $r_1$ & $r_2$ & $t$ & $b_1$ & $b_2$ & $b_3$ & $b'_4$ & $b_5$ & $i$ & $3$ & $g^2_3$
\\
$g_4$ &  $a$ & $-r_1$ & $r_2$ & $-t$ & $b_1$ & $b_2$ & $b_3$ & $b_4$ & $b_5$ & $i$ & $2$ & $g_4 g_5$
\\
$g_5$ & $a$ & $r_1$ & $r_2$ & $t$ & $b_1$ & $b_2$ & $b_3$ & $b_4$ & $-b_5$ & $i$ & $2$ & $g_5$
\\
$\bar{g}_1$ & $a$ & $r_1$ & $r_2$ & $t$ & $b_1$ & $b_2$ & $b_3$ & $b_4$ & $b_5$ & $-i$ & $2$ & $g_5\bar{g}_1$
\\
\bottomrule
\end{tabular}
\hspace{1 cm}
\parbox[t]{8 cm}{ \vspace{0.1 cm} 
$t' =\cos \frac{23\pi}{30} $
\\ 
\vspace{0.05 cm}

$b'_1=-\frac{1}{2}\sqrt{4(-5r_1r_2 -(4a + 23)r_1)t + 2((4a + 21)r_2 +8a + 111)}$
\\ 
\vspace{0.05 cm}

$b''_1=\frac{1}{2}\sqrt{4(5r_1r_2 +(4a - 23)r_1)t' + 2((4a -21)r_2 -8a + 111)}$
\\ 
\vspace{0.05 cm}

$b'_2=2\re\Bigl( e^{\frac{2\pi i}{3}} (1+2i\sqrt{31})^{\frac{1}{3}}\Bigr)$
\\ 
\vspace{0.05 cm}

$b'_3=2\re\Bigl( e^{\frac{2\pi i}{3}} \Bigl(4+b'_2+i\sqrt{48-8b^{\prime}_2-b^{\prime 2}_2}\Bigr)^{\frac{1}{3}}\Bigr)$
\\ 
\vspace{0.05 cm}

$b'_4= 2 \re\Bigl(e^{\frac{2\pi i}{3}} (70+10 i\sqrt{31})^{\frac{1}{3}}\Bigr)$
\\ 
\vspace{0.05 cm}

$b'_5=-\frac{1}{2}\sqrt{2a-18}$

}
}

\vspace{0.2 cm }

\noindent \parbox[t][4 pt][b]{0.3 cm}{2.}
\fbox{\begin{tabular}[t]{ccc}
\toprule
orbit & $\gal(\fd{E}/\fd{Q})$ &  $\gal(\fd{E}/\fd{E}_0)$ 
\\
\midrule \addlinespace[2 pt]
$30abc$ & $\la g_a,g_1,g_2,g_3,g_4,g_5,\bar{g}_1\ra$ & $\la g_1^3,g\vpu{2}_2,g\vpu{2}_3,g\vpu{2}_4,g\vpu{2}_5,\bar{g}\vpu{2}_1\ra$ 
\\ \addlinespace[2 pt]
$30d$ & $\la g_a,g_1,g_2,g_3,g_4,g_5,\bar{g}_1\ra$ & $\la g_1, g_2,g_3,g_4,g_5,\bar{g}_1\ra$
\\ \addlinespace[2 pt]
\bottomrule
\end{tabular}
\hspace{1 cm}
\parbox[t][1  cm][c]{10.75 cm}{ $g_1$ cycles  orbits $30abc$ in the order $30a\to 30b\to 30c \to 30a$, and restricts to an order $3$ automorphism on $30d$.}
}

\vspace{0.5 cm }

\noindent \parbox[t][4 pt][b]{0.3 cm}{3.}
\fbox{\begin{tabular}[t]{ccccccc}
\toprule
$G_{a1}$ & $G_{b1}$ & $G_{c1}$ & $G_{d1}$ & $G_{a2}=G_{c2}$ & $G_{b2}$& $G_{d2}$ 
\\ \addlinespace[2 pt]
\midrule
$\bmt 15 & 49 \\ 4 & 45 \emt$ & $\bmt 9 & 56 \\ 5 & 51 \emt$ & $\bmt 21 & 40 \\ 20 & 1\emt$ & $\bmt 1 & 20\\ 0 & 1\emt$&
$\bmt 13 & 24 \\ 36 & 37\emt$ & $\bmt 7 & 36 \\ 24 & 43 \emt$ & $\bmt 1 & 21\\ 3 & 40 \emt$
\\ \addlinespace[4 pt]
\toprule 
$G_{a3}=G_{c3}$ & $G_{b3}$& $G_{d3}$& $G_{a4}=G_{b4}=G_{c4}$& $G_{d4}$ & $G_{a5}=G_{b5}=G_{c5}$& $G_{d5}$
\\ \addlinespace[2 pt]
\midrule
$\bmt 11 & 25 \\ 35 & 36 \emt$ & $\bmt 5 & 19 \\ 41 & 24 \emt$ & $\bmt 1 & 48 \\ 24 & 13 \emt$ &
$\bmt 1 & 30 \\ 30 & 31 \emt$ & $\bmt 1 & 3 \\ 9 & 58\emt$ & $\bmt 0 & 31 \\ 29 & 31\emt$ & $\bmt 29 &0 \\ 0 & 29 \emt$
\\ 
\bottomrule
\end{tabular}
}

\vspace{0.5 cm}

\noindent \textbf{Dimension 35}

\vspace{0.1 cm }

\noindent
\parbox[t][4 pt][b]{0.3 cm}{1.}
\fbox{\begin{tabular}[t]{cccccccccccccc}
\toprule 
$h$ & $a$ & $r_1$ & $r_2$  & $r_3$ & $b_1$ & $b_2$ & $b_3$ & $b_4$ &  $b_5$ & $b_6$ & $i$ & order & $g\vpu{-1}_a h g^{-1}_a$
 \\
\midrule
$g_a$ & $-a$ & $r_1$ & $r_2$  & $r_3$ & $b'_1$ & $b_2$ & $b_3$ & $b_4$ &  $b_5$ & $b'_6$ & $i$ & $2$ & $g_a$
\\
$g_1$  & $a$ & $r_1$ & $r_2$  & $-r_3$ & $b_1$ & $b_2$ & $b_3$ & $b_4$ &  $b'_5$ & $b_6$ & $i$ & $4$ & $g^3_1$
\\
$g_2$ &  $a$ & $-r_1$ & $r_2$  & $r_3$ & $b_1$ & $b_2$ & $b'_3$ & $b'_4$ &  $b_5$ & $b''_6$ & $i$ & $24$ & $g^{13}_2$
\\
$g_3$ &  $a$ & $r_1$ & $r_2$  & $r_3$ & $-b_1$ & $b'_2$ & $b_3$ & $b_4$ &  $b_5$ & $b_6$ & $i$ & $6$ & $g^5_3$
\\
$g_4$ &  $a$ & $r_1$ & $-r_2$  & $r_3$ & $b_1$ & $b_2$ & $-b_3$ & $b_4$ &  $b_5$ & $b_6$ & $i$ & $2$ & $g^3_3g\vpu{3}_4$
\\
$\bar{g}_1$ & $a$ & $r_1$ & $r_2$  & $r_3$ & $b_1$ & $b_2$ & $b_3$ & $b_4$ &  $b_5$ & $b_6$ & $-i$ & $2$ & $g_c=g^{12}_2 g^3_2 \bar{g}\vpu{3}_1$
\\
\bottomrule
\end{tabular}
\hspace{1 cm}
\parbox[t]{6 cm}{ \vspace{0.1 cm} 
$b'_1=\sqrt{2a-1}$
\\ 
\vspace{0.05 cm}

$b'_2=2 \re \Bigr( e^{\frac{2\pi i}{3}} \bigl(280+210i\sqrt{6}\bigr)^{\frac{1}{3}}\Bigr)$
\\ 
\vspace{0.05 cm}

$b'_3=\sqrt{14(5-r_1)}$
\\ 
\vspace{0.05 cm}

$b'_4=\re\Bigl( e^{\frac{2 \pi i}{3}} \bigl( 28 + 84 i \sqrt{3}\bigr)^{\frac{1}{3}}\Bigr)$
\\ 
\vspace{0.05 cm}

$b'_5=\sqrt{3+r_3}$
\\ 
\vspace{0.05 cm}

$b'_6=-\frac{1}{7}\sqrt{245a+7 r_2 b_3 -49ar_1}$
\\ 
\vspace{0.05 cm}

$b''_6=\frac{i}{7}\sqrt{245a-7 r_2 b'_3 +49ar_1}$

}
}

\vspace{0.2 cm }

\noindent \parbox[t][4 pt][b]{0.3 cm}{2.}
\fbox{\begin{tabular}[t]{ccc}
\toprule
orbit & $\gal(\fd{E}/\fd{Q})$ &  $\gal(\fd{E}/\fd{E}_0)$ 
\\
\midrule \addlinespace[2 pt]
$35bcdg$ & $\la g_a,g_1,g_2,g_3,g_4,\bar{g}_1\ra$ & $\la g_2,g_3,g_4,\bar{g}_1\ra$ 
\\ \addlinespace[2 pt]
$35af$ & $\la g_a,g_1,g_2,g_3,g_4,\bar{g}_1\ra$ & $\la g_2,g_3,g_4,\bar{g}_1\ra$ 
\\ \addlinespace[2 pt]
$35e$ & $\la g_a,g_2,g_3,g_4,\bar{g}_1\ra$ & $\la g_2,g_3,g_4,\bar{g}_1\ra$ 
\\ \addlinespace[2 pt]
$35h$ & $\la g_a,g_2,g_3,g_4,\bar{g}_1\ra$ & $\la g_2,g_3,g_4,\bar{g}_1\ra$ 
\\ \addlinespace[2 pt]
$35i$ & $\la g_a,g_2,g_3,g_4,\bar{g}_1\ra$ & $\la g_2,g_3,g_4,\bar{g}_1\ra$ 
\\ \addlinespace[2 pt]
$35j$ & $\la g_a,g_2,g_3,g_4,\bar{g}_1\ra$ & $\la g_2,g_3,g_4,\bar{g}_1\ra$ 
\\ \addlinespace[2 pt]
\bottomrule
\end{tabular}
\hspace{1 cm}
\parbox[t][1  cm][c]{10.75 cm}{ $g_1$ cycles  orbits $35bcdg$  in the order $35b\to 35c\to 35d \to 35g \to 35b$ and interchanges orbits $35af$; it satisfies the relations $g_1^2 = g_2^{12}$ on $35af$,  $g_1 = g_2^6 g\vpu{6}_4 \bar{g}\vpu{6}_1$ on $35e$; it restricts to the identity on $35h$, $35i$, $35j$.  $g_2$ satisfies the relations $g^{12}_2 = e$ on $35i$  (where $e$ is the identity) and $g^6_2 = g\vpu{6}_4 \bar{g}\vpu{6}_1$ on $35j$.}
}

\vspace{0.5 cm }

\noindent \parbox[t][4 pt][b]{0.3 cm}{3.}
\fbox{\begin{tabular}[t]{ccccccccc}
\toprule
$G_{a1}$ & $G_{b1}$ & $G_{c1}$ & $G_{d1}$ &  $G_{f1}$& $G_{g1}$ & \parbox[t]{2 cm}{ \centering $G_{a2}=G_{b2}=G_{d2}=G_{i2}$ }& \parbox[t]{2 cm}{ \centering $G_{c2}=G_{e2}=G_{f2}=G_{g2}=G_{h2}$  }
\\ \addlinespace[2 pt]
\midrule
$\bmt 15 & 34 \\ 14 & 20 \emt$ & $\bmt 3 & 15 \\ 18  & 32 \emt$ & $\bmt 8 & 10 \\ 18 & 27\emt$ & $\bmt 8 &17\\ 25 & 27\emt$&
 $\bmt  6 & 14 \\ 20 & 29  \emt$ & $\bmt  3 & 22 \\ 25 & 32 \emt$ & $\bmt 2 & 21 \\14 & 23 \emt$ & $\bmt 14 & 33 \\ 2 & 12\emt$
\\ \addlinespace[4 pt]
\toprule 
$G_{j2}$ & \parbox[t]{2 cm}{ \centering $G_{a3}=G_{b3}=G_{d3}$} &  \parbox[t]{2 cm}{ \centering $G_{c3} = G_{e3}=G_{f3}=G_{g3}=G_{h3}=G_{i3}$ } & $G_{j3}$
& \parbox[t]{2 cm}{ \centering $G_{a4}=G_{b4}=G_{d4}$} &  \parbox[t]{2 cm}{ \centering $G_{c4} = G_{e4}=G_{f4}=G_{g4}=G_{h4}$ } & $G_{i4}$ & $G_{j4}$
\\ \addlinespace[2 pt]
\midrule
$\bmt 1 & 33 \\ 2 & 34 \emt$ & $\bmt 15 & 34 \\ 1 & 14 \emt$  & $\bmt 1 & 20 \\ 15 & 21 \emt$ & $\bmt 3 & 22 \\ 13 & 25\emt$
& $\bmt 4 & 11 \\ 24 & 15\emt$ & $\bmt 15 & 24 \\ 11 & 4 \emt$ & $\bmt 15 & 31 \\ 4 & 11\emt$ & $\bmt 0 & 27 \\ 8 & 27 \emt$
\\ \addlinespace[4 pt]
\bottomrule
\end{tabular}
}

\vspace{0.5 cm}

\noindent \textbf{Dimension 39}

\vspace{0.1 cm }

\noindent
\parbox[t][4 pt][b]{0.3 cm}{1.}
\fbox{\begin{tabular}[t]{cccccccccccccc}
\toprule 
$h$ & $a$ & $r_1$ & $r_2$  &  $b_1$ & $b_2$ & $b_3$ & $b_4$ &  $b_5$ & $b_6$ & $b_7$ & $i r_3$ & order & $g\vpu{-1}_a h g^{-1}_a$
 \\
\midrule
$g_a$ & -$a$ & $r_1$ & $r_2$  &  $b_1$ & $b_2$ & $b'_3$ & $b'_4$ &  $b_5$ & $b'_6$ & $b'_7$ & $i r_3$ & $2$ & $g_a$
\\
$g_1$  & $a$ & $-r_1$ & $r_2$  &  $b'_1$ & $b_2$ & $b''_3$ & $b''_4$ &  $b_5$ & $b_6$ & $b_7$ & $i r_3$ & $8$ & $g^7_1$
\\
$g_2$ &  $a$ & $r_1$ & $r_2$  &  $-b_1$ & $b_2$ & $b_3$ & $b_4$ &  $b_5$ & $b_6$ & $b_7$ & $i r_3$ & $2$ & $g_2$
\\
$g_3$ &  $a$ & $r_1$ & $-r_2$  &  $b_1$ & $b'_2$ & $b_3$ & $b_4$ &  $b_5$ & $b_6$ & $b_7$ & $i r_3$ & $4$ & $g^6_1 g\vpu{6}_3 g\vpu{6}_6$
\\
$g_4$ &   $a$ & $r_1$ & $r_2$  &  $b_1$ & $b_2$ & $b_3$ & $b_4$ &  $b'_5$ & $b_6$ & $b_7$ & $i r_3$ & $3$ & $g\vpu{2}_4 g^2_5$
\\
$g_5$ & $a$ & $r_1$ & $r_2$  &  $b_1$ & $b_2$ & $b_3$ & $b_4$ &  $b_5$ & $b''_6$ & $b_7$ & $i r_3$ & $3$ & $g^2_5$
\\
$g_6$ &  $a$ & $r_1$ & $r_2$  &  $b_1$ & $b_2$ & $b_3$ & $b_4$ &  $b_5$ & $b_6$ & $-b_7$ & $i r_3$ & $2$ & $g_6$
\\
$\bar{g}_1$ & $a$ & $r_1$ & $r_2$  &  $b_1$ & $b_2$ & $b_3$ & $b_4$ &  $b_5$ & $b_6$ & $b_7$ & $-i r_3$ & $2$ & $g_c = g_6 \bar{g}_1$
\\
\bottomrule
\end{tabular}
\hspace{1 cm}
\parbox[t]{7 cm}{ \vspace{0.1 cm} 
$b'_1 = \sqrt{-3 r_1+18}$
\\
\vspace{0.05 cm}

$b'_2=-\sqrt{-18r_2+78}$
\\
\vspace{0.05 cm}

$b'_3=\sqrt{(-4a+15)(r_1+5)}$
\\ 
\vspace{0.05 cm}

$b''_3=-\sqrt{(4a+15)(-r_1+5)}$
\\ 
\vspace{0.05 cm}

$b'_4=-\sqrt{-(2a+5)(r_1-2)b'_3 +(8a+35)(10-r_1)}$
\\ 
\vspace{0.05 cm}

$b''_4=\sqrt{-(2a-5)(r_1+2)b''_3 -(8a-35)(10+r_1)}$
\\ 
\vspace{0.05 cm}

$b'_5=\re\Bigl(e^{\frac{2\pi i}{3}}\bigl(-676+10140i\sqrt{3}\bigr)^{\frac{1}{3}}\Bigr)$
\\ 
\vspace{0.05 cm}

$b'_6=\re\Bigl(\bigl(180+4a+4i\sqrt{6753-90a}\bigr)^{\frac{1}{3}}\Bigr)$
\\ 
\vspace{0.05 cm}

$b''_6=\re\Bigl(e^{\frac{2\pi i}{3}}\bigl(180-4a+4i\sqrt{6753+90a}\bigr)^{\frac{1}{3}}\Bigr)$
\\ 
\vspace{0.05 cm}

$b'_7=\sqrt{6a+3}$

}
}

\vspace{0.2 cm }

\noindent \parbox[t][4 pt][b]{0.3 cm}{2.}
\fbox{\begin{tabular}[t]{ccc}
\toprule
orbit & $\gal(\fd{E}/\fd{Q})$ &  $\gal(\fd{E}/\fd{E}_0)$ 
\\
\midrule \addlinespace[2 pt]
$39acde$ & $\la g_a,g_1,g_2,g_3,g_4,g_5,g_6,\bar{g}_1\ra$ & $\la g^2_1,g\vpu{2}_3,g\vpu{2}_4,g\vpu{2}_5,g\vpu{2}_6,\bar{g}\vpu{2}_1\ra$ 
\\ \addlinespace[2 pt]
$39bf$ & $\la g_a,g_1,g_3,g_4,g_5,g_6,\bar{g}_1\ra$ & $\la g^2_1,g\vpu{2}_3,g\vpu{2}_4,g\vpu{2}_5,g\vpu{2}_6,\bar{g}\vpu{2}_1\ra$ 
\\ \addlinespace[2 pt]
$39gh$ & $\la g_a,g_1,g_2,g_3,g_4,g_6,\bar{g}_1\ra$ & $\la g^2_1,g\vpu{2}_2,g\vpu{2}_3,g\vpu{2}_4,g\vpu{2}_6,\bar{g}\vpu{2}_1\ra$ 
\\ \addlinespace[2 pt]
$39ij$ & $\la g_a,g_1,g_3,g_4,g_6\bar{g}_1\ra$ & $\la g^2_1,g\vpu{2}_3,g\vpu{2}_4,g\vpu{2}_6,\bar{g}\vpu{2}_1\ra$ 
\\ 
\bottomrule
\end{tabular}
\hspace{1 cm}
\parbox[t][1  cm][c]{10 cm}{ $g_1$ interchanges the pairs $39ae$, $39cd$,  $39bf$,  $39gh$ and $39ij$.  $g_2$ interchanges the pairs $39ac$ and  $39de$ and restricts to the identity on $39bf$ and $39ij$.  $g_5$ restricts to the identity on $39gh$ and $39ij$.}
}

\vspace{0.5 cm }

\noindent \parbox[t][4 pt][b]{0.3 cm}{3.}
\fbox{\begin{tabular}[t]{cccccccccc}
\toprule
$G_{a1}$ & $G_{b1}$ & $G_{c1}$ & $G_{d1}$ &  $G_{e1}$& $G_{f1}$ & $G_{g1}$ & $G_{h1}$ & $G_{i1}$ & $G_{j1}$
\\ \addlinespace[2 pt]
\midrule
$\bmt 12 & 14 \\ 25 & 26 \emt$ & $\bmt 0 & 25 \\ 14 & 25 \emt$ &$\bmt 10 & 32 \\ 3 & 29 \emt$ &$\bmt13 & 27 \\ 1 & 26  \emt$ &$\bmt 3 & 7 \\ 32 & 10 \emt$ &
$\bmt 16 & 33 \\ 6 & 10 \emt$ &$\bmt 4 & 30 \\ 29 & 13 \emt$ &$\bmt 2 & 3 \\ 38 & 38 \emt$ &$\bmt 0 & 7 \\ 28 & 6  \emt$ &$\bmt 4 & 21\\ 6 & 22\emt$ 
\\ \addlinespace[4 pt]
\toprule 
$G_{a2}$ & $G_{c2}$ & $G_{d2}=G_{e2}$ & $G_{g2}$ & $G_{h2}$ 
& \parbox[t]{2 cm}{ \centering $G_{a3}=G_{b3}=G_{c3}=G_{e3}=G_{f3}$} 
&$G_{d3}$ & $G_{g3}=G_{h3}$ & $G_{i3}=G_{j3}$
\\ \addlinespace[2 pt]
\midrule
$\bmt 19 & 23 \\ 16 & 3 \emt$ & $\bmt 3 & 16 \\ 23 & 19\emt$ & $\bmt 1 & 12 \\ 13 & 38\emt$ & $\bmt 7 & 7 \\ 15 & 26\emt$ & $\bmt 20 & 7 \\ 15 & 13\emt$
& $\bmt 21 & 25 \\ 14 & 7 \emt$ & $\bmt 7 & 14 \\ 25 & 21\emt$ & $\bmt 11 & 18 \\ 33 & 32 \emt$ & $\bmt 9 & 28 \\ 34 & 33 \emt$ 
\\  \addlinespace[4 pt]
\toprule
\parbox[t]{2 cm}{\centering $G_{a4}=G_{b4}=G_{c4}=G_{e4}=G_{f4}$} 
& $G_{d4}$ & $G_{g4}=G_{h4}$ & $G_{i4}=G_{j4}$ &
\parbox[t]{2 cm}{\centering $G_{a5}=G_{b5}=G_{c5}=G_{e5}=G_{f5}$} & $G_{d5}$
& \parbox[t]{2 cm}{\centering $G_{a6}=G_{b6}=G_{c6}=G_{d6}=G_{e6}=G_{f6}$} 
& $G_{g6}=G_{h6}$ & $G_{i6}=G_{j6}$
\\ \addlinespace[2 pt]
\midrule
$\bmt 4 & 9 \\ 30 & 13 \emt$  &  $\bmt 13 & 30 \\ 9 & 4  \emt$  &  $\bmt 22 & 0 \\ 0 & 22 \emt$  &  $\bmt 0 & 2 \\ 8 & 24 \emt$  
&$\bmt 12 & 14 \\ 25 & 26\emt$ & $\bmt 25 & 27 \\ 12 & 13\emt$
& $\bmt 0 & 1 \\ 38 & 1\emt$ & $\bmt 2 & 3 \\ 12 & 38\emt$ & $\bmt 0 & 7 \\ 28 & 6\emt$
\\
\bottomrule
\end{tabular}
}

\vspace{1 cm}

\noindent \textbf{Dimension 48}

\vspace{0.1 cm }

\noindent
\parbox[t][4 pt][b]{0.3 cm}{1.}
\fbox{\begin{tabular}[t]{ccccccccccccccc}
\toprule 
$h$ & $a$ & $r_1$ & $r_2$  & $r_3$ & $t$ &  $b_1$ & $b_2$ & $b_3$ & $b_4$ &  $b_5$ & $b_6$ &  $i$ & order & $g\vpu{-1}_a h g^{-1}_a$
 \\
\midrule
$g_a$ & $-a$ & $r_1$ & $r_2$  & $r_3$ & $t$ &  $b'_1$ & $b_2$ & $b_3$ & $b_4$ &  $b'_5$ & $b_6$ &  $i$ & $2$ & $g_a$
\\
$g_1$  & $a$ & $r_1$ & $r_2$  & $-r_3$ & $t$ &  $b_1$ & $b_2$ & $b_3$ & $b'_4$ &  $b''_5$ & $b_6$ &  $i$ & $8$ & $g^7_1$
\\
$g_2$ &  $a$ & $-r_1$ & $r_2$  & $r_3$ & $t'$ &  $b_1$ & $b'_2$ & $b'_3$ & $b_4$ &  $b_5$ & $b_6$ &  $i$ & $8$ & $g\vpu{6}_2g^6_3$
\\
$g_3$ & $a$ & $r_1$ & $-r_2$  & $r_3$ & $t''$ &  $b_1$ & $b'_2$ & $b''_3$ & $b_4$ &  $b_5$ & $b_6$ &  $i$ & $8$ & $g^4_1g^3_3$
\\
$g_4$ &  $a$ & $r_1$ & $r_2$  & $r_3$ & $t$ &  $b_1$ & $b_2$ & $b_3$ & $b_4$ &  $b_5$ & $b'_6$ &  $i$ & $3$ & $g^2_4$
\\
$g_5$ & $a$ & $r_1$ & $r_2$  & $r_3$ & $t$ &  $-b_1$ & $b_2$ & $b_3$ & $b_4$ &  $b_5$ & $b_6$ &  $i$ & $2$ & $g_5$
\\
$\bar{g}_1$ & $a$ & $r_1$ & $r_2$  & $r_3$ & $t$ &  $b_1$ & $b_2$ & $b_3$ & $b_4$ &  $b_5$ & $b_6$ &  $-i$  & $2$ & $g_c = g_5 \bar{g}_1$ 
\\
\bottomrule
\end{tabular}
\hspace{1 cm}
\parbox[t]{8 cm}{ \vspace{0.1 cm} 

$t'=\cos\frac{37\pi}{48}$
\\
\vspace{0.05 cm}

$t'' = \cos \frac{17 \pi }{48}$
\\
\vspace{0.05 cm}

$b'_1 =\sqrt{a+1} $
\\
\vspace{0.05 cm}

$b'_2 = -\sqrt{6-r_1r_2}$
\\
\vspace{0.05 cm}

$b'_3 = \sqrt{ 6+2r_1r_2+(r_2+r_1)b'_2 }$
\\
\vspace{0.05 cm}

$b''_3 = -\sqrt{6+2r_1r_2-(r_2+r_1)b'_2}$
\\
\vspace{0.05 cm}

$b'_4 = - \sqrt{-2r_3+42}$
\\
\vspace{0.05 cm}

$b'_5=-\sqrt{(-5(a-1)+(3a+5)r_3)b_4+2(a-5)r_3+210(3+a)}$
\\
\vspace{0.05 cm}

$b''_5= - \sqrt{(5(a+1)+(3a-5)r_3)b'_4+2(a+5)r_3+210(3-a)}$
\\
\vspace{0.05 cm}

$b'_6=2\re\Bigr( e^{\frac{2\pi i}{3}} \bigl( 7 + i \sqrt{15}\bigr)^{\frac{1}{3}}\Bigr)$
}
}

\vspace{0.2 cm }

\noindent \parbox[t][4 pt][b]{0.3 cm}{2.}
\fbox{\begin{tabular}[t]{ccc}
\toprule
orbit & $\gal(\fd{E}/\fd{Q})$ &  $\gal(\fd{E}/\fd{E}_0)$ 
\\
\midrule \addlinespace[2 pt]
$48abcd$ & $\la g_a,g_1,g_2,g_3,g_4,g_5,\bar{g}_1\ra$ & $\la g^4_1,g\vpu{2}_2,g\vpu{2}_3,g\vpu{2}_4,g\vpu{2}_5,\bar{g}\vpu{2}_1\ra$ 
\\ \addlinespace[2 pt]
$48e$ & $\la g_a,g_1,g_2,g_3, g_5,\bar{g}_1\ra$ & $\la g_1,g_2,g_3,g_5,\bar{g}_1\ra$ 
\\ \addlinespace[2 pt]
$48f$ & $\la  g_a,g_2,g_3,g_4,g_5,\bar{g}_1\ra$ & $\la g_2,g_3,g_4,g_5,\bar{g}_1\ra$ 
\\ \addlinespace[2 pt]
$48g$ & $\la g_a,g_2,g_3,g_5,\bar{g}_1\ra$ & $\la g_2,g_3,g_5,\bar{g}_1\ra$ 
\\ 
\bottomrule
\end{tabular}
\hspace{1 cm}
\parbox[t][1  cm][c]{10 cm}{ $g_1$ cycles $48abcd$ in the order $48a \to 48d \to 48b \to 48c \to 48a$ and restricts to the identity on $48f$, $48g$.  $g_4$ restricts to the identity on $48e$, $48g$.   }
}

\vspace{0.5 cm }

\noindent \parbox[t][4 pt][b]{0.3 cm}{3.}
\fbox{\begin{tabular}[t]{cccccccccc}
\toprule
$G_{a1}$ & $G_{b1}$ & $G_{c1}$ & $G_{d1}$ &  $G_{e1}$ 
& \parbox[t]{2 cm}{\centering $G_{a2}=G_{b2}=G_{f2}$} 
& $G_{c2}=G_{d2}$ & $G_{e2}$ & $G_{g2}$
\\ \addlinespace[2 pt]
\midrule
$\bmt 13 & 56 \\ 69 & 83  \emt$  &  $\bmt  5 & 19 \\ 24 & 91 \emt$  &  $\bmt 8 & 21 \\ 29 & 88 \emt$  &  $\bmt 5 & 72 \\ 77 & 91 \emt$  &  $\bmt 11 & 50 \\ 58 & 89  \emt$  &  
$\bmt1 & 11 \\ 85 & 12 \emt$  &  $\bmt 12 & 85 \\ 11 & 1\emt$  &  $\bmt 23 & 36 \\ 84 & 83 \emt$  & $\bmt 0 & 1 \\ 37 & 95 \emt$  
\\ \addlinespace[4 pt]
\toprule 

\\ \addlinespace[2 pt]
\midrule
\\  \addlinespace[4 pt]
\toprule
 \parbox[t]{2 cm}{\centering $G_{a3}=G_{b3}=G_{f3}$} 
& $G_{c3}=G_{d3}$ & $G_{e3}$ & $G_{g3}$
& \parbox[t]{2 cm}{\centering $G_{a4}=G_{b4}=G_{f4}$} 
& $G_{c4}=G_{d4}$ 
& \parbox[t]{3.5 cm}{\centering $G_{a5}=G_{b5}=G_{c5}=G_{d5}=G_{e5} = G_{f5}=G_{g5}$} 
\\ \addlinespace[2 pt]
\midrule
$\bmt 17 & 37 \\ 59 & 54  \emt$  &  $\bmt  11 & 31 \\ 65& 42  \emt$  &  $\bmt 25 & 93 \\ 81 & 28  \emt$  &  $\bmt 4 & 47 \\ 11 & 53  \emt$  &
$\bmt 15 & 17 \\ 79 & 32  \emt$ &  $\bmt 31 & 33 \\ 63 & 64 \emt$ & $\bmt 95 & 0 \\ 0 & 95 \emt$
\\
\bottomrule
\end{tabular}
}
}
\end{landscape}

\section*{Acknowledgements}
This research was supported in part by the Australian Research Council via EQuS project number CE11001013.
SF acknowledges support from an Australian Research Council Future Fellowship FT130101744.

\end{document}